\documentclass{article}[12pt]
\usepackage{jheppub}
\usepackage[utf8]{inputenc}
\usepackage{amsmath}
\usepackage{amssymb}
\usepackage{amsfonts}
\usepackage{amsthm}
\usepackage{xcolor}
\usepackage{rotating}
\usepackage{array}
\usepackage{appendix}
\usepackage{bbold}
\usepackage{textcomp}

\def\FG{G}
\def\sc{\textit{sc}}
\def\seed{\textit{seed}}
\title{Crossing Symmetry for \\[4mm] Long Multiplets in 4D $\mathcal{N}=1$ SCFTs}

\author{Ilija Buri\' c$^1$,}

\author{Volker Schomerus$^1$}

\author{and Evgeny Sobko$^{2,3}$}

\affiliation{$^1$DESY, Notkestra\ss e 85, D-22607 Hamburg, Germany}

\affiliation{$^2$St. Petersburg Department of V.A. Steklov Mathematical Institute and Euler International Mathematical Institute of the Russian Academy of Sciences}

\affiliation{$^3$School of Physics and Astronomy, University of Southampton,\ Highfield, Southampton, SO17 1BJ, United Kingdom}

\preprint{DESY 20-205}

\abstract{In this work we construct the crossing symmetry equations for mixed
correlators of two long and two BPS operators in 4D $\mathcal{N}=1$ SCFTs. The
analysis presented here illustrates how our general group theoretic approach
to long superblocks and tensor structures of superconformal algebras can be
applied to give explicit ready-to-use expressions. In the case at hand, we
obtain a system of four crossing symmetry equations for the relevant OPE
coefficients. One of these four equations coincides with the equation found
and analysed by Li, Meltzer and Stergiou by restricting to the superprimary
component of the long multiplets. The other three equations are new and they
provide powerful additional constraints on the same OPE data.}

\begin{document}

\maketitle

\section{Introduction}

After its revival in \cite{Rattazzi:2008pe}, the conformal bootstrap has proved to be a powerful
approach to conformal field theories (CFTs) which offers insights, particularly about their landscape,
that is currently unattainable by other methods. Within the bootstrap approach, CFTs are found as
solutions to an infinite set of crossing symmetry equations. In order to write down these equations,
one needs to have a good control over two kinematical ingredients - conformal partial waves and
tensor structures. The former are a natural basis for the expansion of correlation functions in
terms of conformal invariants while the latter capture the nontrivial transformation behaviour
of correlation functions under rotations.

In their pioneering work \cite{Dolan:2000ut,Dolan:2003hv}, Dolan and Osborn realised that four-point
conformal blocks can be characterised as solutions of Casimir differential equations, and used this
fact to find series expansions for them. While these results applied to blocks for correlation functions of
scalars in even number of spacetime dimensions, they have since been extended to spinning operators
and odd dimensions. In most cases, explicit general formulas for the blocks, e.g. in terms of special functions, do not exist, but any
given conformal block can be computed in an algorithmic way by starting from the scalar ones and
using weight-shifting operators. For this and other techniques of finding spinning conformal blocks,
see \cite{Dolan:2011dv,Costa:2011dw,SimmonsDuffin:2012uy,Hogervorst:2013sma,Penedones:2015aga,
Echeverri:2016dun,Karateev:2017jgd,Isachenkov:2017qgn,Erramilli:2019njx,Fortin:2019fvx,
Fortin:2019dnq,Fortin:2020ncr}.

Given that most conformal field theories we know about possess some amount of supersymmetry, in
particular in $d=4$ dimensions, it is certainly of interest to adapt and apply the bootstrap programme
to superconformal field theories (SCFTs) and clearly a lot of research has explored this important
direction already. The superconformal bootstrap programme meets some obvious challenges, namely the
fact that even the smallest supermultiplets include several primaries for the bosonic conformal
symmetry, certainly also including fields with spin. Hence, studying SCFTs as conformal field
theories without taking the additional constraining power of supersymmetry into account seems
a rather daunting task. On the other hand, the kinematical aspects and in particular the partial waves
for fields in generic representations of the superconformal algebra are very poorly developed.
So far, the strategy has been largely to either restrict the bootstrap analysis to four-point
function of chiral (half-BPS) operators or to four-point functions of the superprimary component.
For these two cases, it has been possible to construct the relevant blocks, see e.g.
\cite{Fitzpatrick:2014oza,Khandker:2014mpa,Bissi:2015qoa,Doobary:2015gia,Bobev:2017jhk,Li:2016chh,Lemos:2015awa,
Chang:2017xmr,Liendo:2016ymz,Lemos:2016xke,Liendo:2018ukf,Gimenez-Grau:2019hez,Gimenez-Grau:2020jrx}
for more recent work and references to earlier literature. On the other hand, it was demonstrated in
\cite{Cornagliotto:2017dup} that superconformal bootstrap imposes significantly more
powerful constraints when applied to correlators involving long or at least semi-short operators.
This should not come as a surprise since a long multiplet bootstrap combines the usual power of
a mixed correlator bootstrap with the full constraints of supersymmetry that keep the number of
independent dynamical variables small. In \cite{Cornagliotto:2017dup}, the long multiplet
bootstrap was only explored for $\mathcal{N} = (2,0)$ SCFTs in $d=2$ dimensions since for this
case it was possible to construct all the relevant blocks with rather traditional techniques,
see also \cite{Kos:2018glc} for another instance of the long multiplet bootstrap $2$-dimensional
SCFTs. For promising applications to higher dimensions, our ignorance about generic superconformal
blocks has been the main bottleneck. In fact, with the exception of \cite{Ramirez:2018lpd} there
had been very few attempts even to advance beyond the present status.

In \cite{Buric:2019rms,Buric:2020buk} we launched a programme to resolve this issue of superconformal
partial wave decompositions for long operators, at least for superconformal algebras of type I, for which
the (internal) R-symmetry group $U$ contains an abelian factor $U(1)$. In particular,
all superconformal algebras in $d=4$ are of type I. As usual, the analysis consists of two parts.
Starting with some correlator of fields that are inserted at points $x_i$ one first needs to find
an appropriate set of four-point tensor structures that reduce the correlator to functions of
superconformal cross ratios. Once these tensor structures are found, one derives and solves
Casimir differential equations for a basis of superconformal blocks. We addressed the latter step
in \cite{Buric:2020buk} after we had dealt with the former in \cite{Buric:2019rms}. Both parts
of the construction are heavily based on tools from group theory and harmonic analysis,
following an earlier group theoretic approach to spinning conformal blocks that was
developed in \cite{Schomerus:2016epl,Schomerus:2017eny,Buric:2019dfk}. With the choice of
tensor structures and coordinates this approach provides, the Casimir differential equations
for the blocks assume a universal form of a matrix Schr\"odinger problem of Calogero-Sutherland
type. For this reason we shall refer to the group theoretic coordinates as Calogero-Sutherland
coordinates and to the group theoretic choice of tensor structures as the Calogero-Sutherland
gauge. When applied to superconformal algebras of type I, it turns out that the Casimir
differential equations in the Calogero-Sutherland gauge take a particularly simple form -
they can be written as a nilpotent perturbation of a set of Casimir equations for
spinning fields of the (bosonic) conformal algebra. This implies that one can construct the
superblocks through a finite order perturbation theory from spinning bosonic ones. While the
general framework developed in \cite{Buric:2019rms} is fully algorithmic, it remained a bit
formal. It is the main goal of the present work to illustrate how to apply this formalism in
a concrete example relevant to 4-dimensional SCFTs. Our presentation will focus on the
derivation of the crossing symmetry equations, i.e.\ the compatibility of the $s-$ and 
$t-$channel superconformal partial wave expansions. In these equations, the relevant tensor structures
only enter through the quotient of the $s-$ and $t-$channel quantities. This quotient, which
we dubbed the crossing factor in \cite{Buric:2020buk}, is a matrix of conformal invariants,
i.e.\ it only depends on cross ratios. Therefore, the crossing matrix is a little easier to
compute than the individual tensor structures, see below.
\medskip

The example we will discuss here
is relevant for the study of 4-dimensional $\mathcal{N}=1$ superconformal field theories and concerns correlations functions
\begin{equation}\label{4-point-function}
G_4(x_i) = \langle \bar\varphi_1(x_1) \mathcal{R}(x_2) \varphi_3(x_3) \mathcal{R}(x_4) \rangle,
\end{equation}
of a chiral field $\varphi_3$, an anti-chiral $\bar\varphi_1$ and two identical long multiplets
$\mathcal{R}$ with real scalar superprimary component $R$. For this example we shall derive the full set of long multiplet
crossing equations. Once all the algebraic dust of \cite{Buric:2019rms} and \cite{Buric:2020buk} has
settled, the equations take a fully explicit form that is ready-to-use.

\vskip0.1cm Crossing symmetry for four-point functions of the corresponding superprimary fields has been studied
before, most notably in \cite{Li:2017ddj}. In spelling out our result and comparing it to the equation
in \cite{Li:2017ddj} we will adopt notations of this earlier work, in particular for the dynamical
operator product coefficients in the crossing equations. To understand the main features of the
crossing equations let us briefly recall that the operator product of $\bar \varphi$ and $\mathcal{R}$
contains four different families of supermultiplets with superprimaries $\mathcal{O}$ labelled by their
weights, spins and $U(1)$ internal charges. It is important to stress that 3-point functions of the superprimary field $\mathcal{O}$
with $\bar \varphi$ and $R$ can vanish and that a non-trivial three-point function requires to pass from
$\mathcal{O}$ to one of its superdescendants. In the case at hand, this comment applies to three of
the four families of operators we mentioned above. Following \cite{Li:2017ddj}, we shall denote the
associated three-point couplings by
\begin{equation} \label{eq:opecoeff}
c_{\bar\varphi R (\bar Q^2\mathcal{O})_l}\ , \quad
\hat c_{\bar\varphi R (\bar Q\mathcal{O})_l}\ , \quad
\check c_{\bar\varphi R(\bar Q\mathcal{O})_l} \ , \quad
\bar c_{\bar\varphi R\mathcal{O}_l} \ .
\end{equation}
This notation displays the relevant superdescendant explicitly. Let us point out that the normalisation
of any superdescendant is determined through supersymmetry by the (canonical) normalisation of the
superprimary. With these notations being set up, the crossing equations we are about to derive assume
the following form
\begin{equation} \label{eq:crossing}
   \bar \sum |c_{\bar\varphi R (\bar Q^2\mathcal{O})_l}|^2\ \mathcal{F}_1  +
    \hat \sum |\hat c_{\bar\varphi R (\bar Q\mathcal{O})_l}|^2\,  \hat\gamma\ \mathcal{F}_2
   + \check \sum |\check c_{\bar\varphi R(\bar Q\mathcal{O})_l}|^2 \, \check \gamma\  \mathcal{F}_3
   + \sum |\bar c_{\bar\varphi R\mathcal{O}_l}|^2 \, \bar \gamma\  \mathcal{F}_4 = 0\ .
\end{equation}
The four different terms in this equation correspond four independent four-point tensor structures.
Each of the summations runs over intermediate supermultiplets. The last summation $\bar \Sigma$,
for example, includes all superprimaries $\mathcal{O}^{\Delta}_l$ of any spin $l$ and weight $\Delta$
such that the three-point function with the fields $\bar \varphi$ and $R$ does not vanish etc.

The factors $\gamma$ that appear in three of the terms in eq.\ \eqref{eq:crossing} are some
explicitly known rational functions of the weight $\Delta$ and the spin $l$ that are needed to
match our conventions for the normalisation of superconformal blocks with the ones used in
\cite{Li:2017ddj}, see below. So the main kinematic ingredient in eq.\ \eqref{eq:crossing}
are the objects $\mathcal{F}_j$ which take the form
\begin{equation} \label{eq:Fjdef}
\mathcal{F}_j = \FG_j(\alpha_i^{-1}) - M_{st}(\alpha_i) \FG_j(\alpha_i),
\end{equation}
of a weighted sum of $s-$ and $t-$channel superconformal blocks $\FG_j$, as usual. The arguments $\alpha_i$ are functions of the two cross ratios defined in eq.\ \eqref{eq:alpha} and
the weight factor $M_{st}(\alpha_i)$ is the crossing factor we mentioned before, i.e. the ratio of $s-$ and $t-$channel tensor structures. 
Superconformal blocks $\FG_j$ possess four independent components $\FG_{jk}$, $k=1, \dots,4$. We construct them explicitly in terms of well-known (spinning) bosonic blocks in eqs.\ (\ref{eq:F1})-(\ref{eq:k2}). 
All these partial waves can be expressed in terms of Gauss' hypergeometric function through formulas given in the appendix E. The crossing factor $M_{st}$ is a $4 \times 4$ matrix spelled out in eq.\ \eqref{eq:Mst}. Together, these formulas allow to evaluate the four components of $\mathcal{F}_j$ for any value of the cross ratios. In conclusion, our equation \eqref{eq:crossing} provides a set of four independent crossing equations in which all the
input from kinematics is explicitly known.

In comparison, the work of Li, Meltzer and Stergiou contains only a single such crossing equation for the correlator \eqref{4-point-function} that arises from the superprimary component $R$ of
the long multiplet $\mathcal{R}$, see eq.\ $(4.9)$ of \cite{Li:2017ddj}. As one can check by explicit comparison of the superblocks, this equation coincides with the first of the four
components in our crossing equation \eqref{eq:crossing}. In fact, the partial waves of \cite{Li:2017ddj} can be obtain from the superconformal blocks that we construct in eqs. (\ref{eq:F1})-(\ref{eq:k2}) 
as wavefunctions of a matrix Calogero-Sutherland problem, by restricting to their first components. More precisely
\begin{equation}
F = \Lambda \mathcal{F}_{11}\ , \quad
\hat{\mathcal{F}} = \hat\gamma \Lambda \mathcal{F}_{21} \ , \quad
\check{\mathcal{F}} = \check\gamma \Lambda \mathcal{F}_{31} \ , \quad
\bar{\mathcal{F}} = \bar\gamma \Lambda \mathcal{F}_{41} \ .
\end{equation}
Explicit formulas for the coefficients $\hat \gamma, \check \gamma$ and $\gamma$ are given in eq.\ \eqref{eq:gamma} and the factor $\Lambda$ is a simple function of the cross ratios that maps blocks from Calogero-Sutherland gauge to the more standard gauge used in \cite{Li:2017ddj}, see \cite{Isachenkov:2016gim}. Since this factor is the same for all four blocks, $\Lambda$ does not appear in the crossing equation. Hence our result \eqref{eq:crossing} is fully compatible with the work of Li et al. and extends it, as we obtain three more equations that the same dynamical coefficients have to satisfy. In this sense, the crossing symmetry in the long multiplet bootstrap is significantly more constraining that its restriction to superprimaries.
\smallskip 

Let us now briefly describe the outline of this paper. In section 2 we adapt the formalism developed 
in \cite{Buric:2020buk} to the computation of the crossing factor $M_{st}(\alpha_i)$ that appears in 
eq.\ \eqref{eq:Fjdef}. The section includes a brief review of the main steps that need to be performed 
for the group theoretic construction of crossing factors. Then these steps are carried out for the 
example at hand. For the initial steps we keep the explicit calculations a bit more general so that 
many of the formulas we present still apply to superconformal algebras of the form $\mathfrak{sl}(2m|
\mathcal{N})$. But in the final computation of $M_{st}$ we then focus on the crossing factor for the 
correlator \eqref{4-point-function} in 4-dimensional $\mathcal{N}=1$ SCFTs. 

Section 3 is devoted to the computation of the superconformal blocks $\FG_j,j=1, \dots, 4$. This is 
achieved by adapting the framework of \cite{Buric:2019rms} to the correlator \eqref{4-point-function}. 
In the first subsection, we review the form of the Laplacian on type I supergroups and discuss its 
rather astonishing features. Then we reduce the Laplacian on the supergroup $G=SL(4|1)$ to a subspace 
of so-called $K$-spherical functions on $G$ that is determined by the choice of fields that appear in 
the 4-point function \eqref{4-point-function}. After this reduction we obtain a $4 \times 4$ matrix 
Schr\" odinger operator whose eigenfunctions are superconformal blocks in Calogero-Sutherland gauge. 
These eigenfunctions are found explicitly in the third subsection before all our results are put 
together in the final subsection to derive the crossing equation \eqref{eq:crossing}.  

Our paper ends with an outlook to further applications of methods from \cite{Buric:2019rms,
Buric:2020buk}. A number of appendices contain some Lie-theoretic background and the set of 
conventions used throughout the main text.

\section{Tensor Structures and Crossing Factor}

In this section we compute the crossing factor $M$ for the correlation function
\eqref{4-point-function}. The first subsection contains a brief review that of the
general procedure that was developed in \cite{Buric:2020buk}. Here we shall present
it in the form of a algorithm that runs through three well defined steps of group
theoretic calculations. These are then executed in the remaining subsections. Part
of the necessary formulas are derived for the larger family of superalgebras
$\mathfrak{sl}(2m|\mathcal{N})$ and hence apply beyond the superconformal algebra
of $\mathcal{N}=1$ SCFTs in $d=4$ dimensions.

\subsection{Review of the group theoretic approach}
\def\Lie{\mathit{Lie}}

Within the group theoretic approach, in which the Casimir equations for conformal
blocks take the form of Calogero-Sutherland eigenvalue equations, the derivation
of the crossing symmetry equations relies on two decompositions of the (super)conformal
group, the Bruhat decomposition and the Cartan decomposition. The Bruhat factorisation
is intimately related to conformal Ward identities and can be used to extend conformal
fields to functions on the group. On the other hand, the Cartan factorisation provides
a definition of Calogero-Sutherland (radial) coordinates. The relation between the two
decompositions for the bosonic conformal group $SO(d+1,1)$ is well-known and this allowed
us to compute the crossing factors of spinning bosonic fields in \cite{Buric:2020buk}.
Our goal is to adapt this program to superconformal groups of type I.
\medskip

Given some superconformal group $G$, one may think of superspace $M$ as the subgroup
that is generated translations and supertranslations. Let $\{X_a\}$ be the associated
generators in the Lie superalgebra and denote by $x = \{x^a\}$ the corresponding superspace
coordinates. The supergroup element $m(x) \subset G$ is then given by
\begin{equation}\label{eq:m}
    m(x) = e^{x^a X_a}\ .
\end{equation}
We shall use this map to identify the superspace $M$, i.e. the graded algebra generated
by the superspace coordinates with the associated subgroup $M \subset G$. Given any pair
of superspace coordinates $x_i, x_j$ one can define coordinates $x_{ij}$ in the graded
tensor product $M_i\otimes M_j$ of the individual superspaces,
\begin{equation}
    m(x_{ij}) = m(x_j)^{-1} m(x_i)\ . \label{xij}
\end{equation}
Any superconformal group contains the so-called Weyl inversion. This is an element of the
underlying bosonic Lie group $G_{(0)}$ that is defined by
\begin{equation}\label{Weyl-inversion}
    w = e^{\pi\frac{K_d-P_d}{2}},
\end{equation}
in terms of the generators $P_d$ of translations and $K_d$ of special conformal
transformations. In bosonic theories, $w$ is the composition of the conformal inversion
with the reflection in the hyperplane orthogonal to the unit vector $e_d$. We use the
Weyl inversion to define a second family of supergroup elements $n(x)$ by
\begin{equation}
    w^{-1} m(x^a) w = n(x^a)\ . \label{1}
\end{equation}
Note that the elements $n(x)$ can be written by exponentiating a linear combination
of the generators of (super)special conformal transformations. With this basic notation
set up, we can now state the \textit{first step} of our construction. In mathematical
terms it instructs us to compute the Bruhat decomposition of $w m(x)$,
\begin{equation}
    w m(x) = m(y(x)) n(z(x)) k(t(x))\ . \label{matrix-identity}
\end{equation}
The Bruhat decomposition is used to write an arbitrary supergroup element $g$ as
a product of an element $m(y) \in M$, and element $n(z) \in N = w^{-1} M w$ and
a third element $k$ that lies in the (bosonic) subgroup $K$ that is generated by
rotations, dilations and internal (R-symmetry) transformations. In other words,
it is associated to the decomposition
\begin{equation}
    \mathfrak{g} = \mathfrak{m} \oplus \mathfrak{n} \oplus \mathfrak{k}, \nonumber
\end{equation}
of Lie superalgebra $\mathfrak{g}$ of $G$ into (super)translation, a (super)special
conformal transformation and elements that commute with with the generator of dilations.
By performing the the Bruhat decomposition \eqref{matrix-identity} for the special elements
$g = w m(x)$ we determine three sets of functions $y(x)$, $z(x)$ and $t(x)$ on superspace.
We shall often simply write $y,z,t$ or $y_{ij}, z_{ij}, t_{ij}$ without specific reference
to the dependence on $x$, $x_{ij}$ etc. Let us also note that $y$ simply encodes how the
Weyl inversion acts on superspace.
\medskip

Once we have determined the functions $y,z,t$ by performing the Bruhat decomposition
\eqref{matrix-identity} of $wm$, we can move to the \textit{second step} of our algorithm.
It involves another factorization of (super)group elements $g$, namely the the Cartan
decomposition. In fact, we will make use of two closely related decompositions which we
will refer to as unprimed and primed Cartan decompositions. In these, supergroup elements
are written in factorised form as
\begin{equation}  \label{Cartan-decompositions}
g = k_l \eta'_l a \eta'_r k_r = \eta_l k_l a k_r \eta_r\ .
\end{equation}
Here $k_l, k_r\in K$ and $a$ lies in the two-dimensional abelian $A$ group that is generated
by $\{P_1+K_1,P_2-K_2\}$. We shall parametrise $A$ by local coordinates $(u_1,u_2)$ according to
\begin{equation}\label{adef}
a(u_1,u_2) =  e^{\frac{u_1+u_2}{4}(P_1+K_1) -i \frac{u_1-u_2}{4}(P_2 - K_2)}\ .
\end{equation}
The factors $\eta_{l,r}$ and $\eta'_{l,r}$ are associated with fermionic generators. More
specifically, $\eta_l,\eta'_l$ are obtained by exponentiation of generators of negative
$R$-charge and $\eta_r,\eta'_r$ from generators with positive charge. That factors $k_{l,r}$
and $a$ are the same in the two versions of the Cartan decomposition follows from the
Baker-Campbell-Hausdorff formula. There is a freedom in the choice of $k_{l,r}$ coming
from the fact that $A$ has a non-trivial stabiliser group $B$ in $K$, but the crossing
equations do not depend on this choice, see \cite{Buric:2020buk}.

Having introduced the Cartan decomposition, we can now state the second step in our
construction explicitly. To this end, let us consider four points $x_i, i=1, \dots,4,$
in superspace and assume that we have calculated the functions $y_{ij},\ z_{ij}$ and
$t_{ij}$ from the Bruhat decomposition of $wm(x_{ij})$. We use this data to construct
the following family of supergroup elements
\begin{equation}
    g(x_i) = n(y_{21})^{-1} m(x_{31}) n(y_{43})\ ,
\end{equation}
that depends on the superspace coordinates $x_i$ of our four external points. The main
challenge of step 2 is to perform the Cartan decomposition of $g(x_i)$. We can do that
once and for all but then need to apply these decompositions to different `channels'.
These channels can be labelled by permutations $\sigma$ of the four points with the
$s$-channel associated to the trivial permutation $\sigma=1$ and the $t$-channel to
the permutation $\sigma=(24)$ that exchanges point $x_2$ and $x_4$. The corresponding
Cartan decompositions read
\begin{equation}\label{Cartan-factors}
 g_\sigma(x_i) = g(x_{\sigma(i)}) = k_{\sigma,l}(x_i) \eta'_{\sigma,l}(x_i)
 a_\sigma(x_i)\eta'_{\sigma,r}(x_i) k_{\sigma,r}(x_i) \ .
\end{equation}
Finding explicit formulas for the various factors requires some work even for bosonic
theories \cite{Buric:2019dfk}. But depending on the precise setup, some shortcuts are
possible, see below.
\medskip

Having constructed the Cartan decompositions \eqref{Cartan-factors} we now turn to the
\textit{third and final step}, namely the construction of the crossing factor for the
transformation from the $s-$ to the $t-$channel. By definition, the crossing factor is
a matrix that acts on the finite dimensional space of polarizations of the external fields.
It may be regarded as a quotient of $s-$ and $t-$channel tensor structures. Let us
suppose that the superprimaries of our four supermultiplets transform in representations
$\rho_i$ of the bosonic subgroup $K$. The choice of representations $\rho_i$ amounts to
the choice of conformal weights $\Delta_i$, spins $\lambda_i$ and $R$-charges $r_i$. The
associated carrier spaces $V_i$ are spanned by the polarization vectors of the
superprimaries. For scalar superprimaries in theories with abelian $R$-symmetry, for
example, the representations $\rho_i$ are 1-dimensional. Given any choice $\rho_i$ of
representations, we define the {\it super-crossing factor} as
\begin{equation}\label{super-crossing-factor}
    \mathcal{M}_{st} = \mathcal{P}_t \Big( \rho_1(\kappa_1)\otimes\rho_2(\kappa_2)
    \otimes\rho_3(\kappa_3)\otimes\rho_4(\kappa_4) \Big)\mathcal{P}_s\ .
\end{equation}
In this formula $\mathcal{P}_s$ and $\mathcal{P}_t$ denote projectors to the subspace of
polarizations that are invariant under the action of the subgroup $B$ that stabilises
four points, see \cite{Buric:2020buk} for details. More importantly, the arguments
$\kappa_i \in K \subset G$ are constructed as
\begin{align}\label{kappa-i}
& \kappa_1 = k_{t,l}^{-1}k_{s,l}, \quad
  \kappa_2 = k^{w}_{t,r} k(t_{23})k(t_{21})^{-1} k^{w}_{s,l}, \\[2mm]
& \kappa_3 = k_{t,r}k_{s,r}^{-1}, \quad
  \kappa_4 = (k^{w}_{t,l})^{-1} k(t_{41})k(t_{43})^{-1} (k^{w}_{s,r})^{-1}, \label{kappa-ii}
\end{align}
from factors that have been determined in previous steps and $k^w = w k w^{-1}$. By construction
$\mathcal{M}$ is a
family of matrices that acts in the tensor product $\otimes_{i=1}^4 V_i$ of the polarization spaces. The
matrix elements are functions of the four sets of superspace coordinates $x_i$. As one
might suspect and we proved in \cite{Buric:2020buk}, the crossing factor $\mathcal{M}_{st}$
is a superconformal invariant, i.e. it depends on the external points $x_i$ only through
their cross ratios. But as it stands, these still include fermionic (nilpotent) cross
ratios in addition to the pair of bosonic cross ratios. To capture the contribution of
all the components of the supermultiplets to the crossing equations one finally has to
perform a Taylor expansion in the fermionic cross ratios to construct the crossing
factor $M_{st}$ that now acts on the space of $B$-invariant polarisations of the
supermultiplet, not just its superprimaries.
\medskip

Let us summarize once again the key steps of our construction. In order to derive the
crossing factor, one needs to determine two factorisations \eqref{matrix-identity} and
\eqref{Cartan-factors}. The results are then substituted in eqs. \eqref{kappa-i},
\eqref{kappa-ii} and \eqref{super-crossing-factor} to find the super-crossing factor.
A final expansion in fermionic invariants then gives the crossing factor $M$ that
appears in the crossing equations. In concrete implementations one can often shortcut
the full Cartan decomposition. Using conformal invariance of the crossing factor it
is usually possible to make a particular gauge choice for the set of $x_i$ that does
not effect the values of cross ratios but significantly simplifies the Cartan decomposition of
$g_\sigma(x_i)$.

We can now detail the content of the remaining subsections. In section 2.2 we obtain
the Bruhat decomposition \eqref{matrix-identity} for two infinite families of type I
superconformal groups, $SL(2|\mathcal{N})$ and $SL(4|\mathcal{N})$. Next in section 2.3
we find the Cartan factors \eqref{Cartan-factors} associated with the subgroup $A$,
dilations and $U(1)_R$-symmetries. These factors are sufficient to write down the
super-crossing factor if the fields in the correlation function are scalars, which is
also done at the end of subsection 2.3. Though explicit, the resulting expression
for the crossing factor is still somewhat complicated. In the last subsection, we
specialise the discussion to the case of two long and two short operators in an
$\mathcal{N}=1$ SCFT, for which the formulas simplify significantly. Finally,
we will perform the remaining Taylor expansion in the fermionic invariants and write
the crossing factor $M_{st}$, thereby completing the task we had set out for this
section.

\subsection{Bruhat decomposition for $\mathfrak{sl}(2m|\mathcal{N})$}

Let $\mathfrak{g}$ be a simple complex superconformal algebra of type I. These include
a few isolated Lie superalgebras, namely $\mathfrak{osp}(2|4),\mathfrak{psl}(2|2)$ and
$\mathfrak{psl}(4|4)$, as well as two infinite families $\mathfrak{sl}(2|\mathcal{N})$
and $\mathfrak{sl}(4|\mathcal{N})$. We shall focus on these infinite families, i.e. we
set $\mathfrak{g} = \mathfrak{sl}(2m|\mathcal{N})$. For this choice of $\mathfrak{g}$
one finds
\begin{equation}
    \mathfrak{g}_{(0)} = \mathfrak{sl}(2m) \oplus \mathfrak{sl}(\mathcal{N}) \oplus
    \mathfrak{u}(1),\quad \mathfrak{g}_{+} = (2m,\overline{\mathcal{N}},1),\quad \mathfrak{g}_- = (\overline{2m},\mathcal{N},-1)\ .
\end{equation}
The summand $\mathfrak{sl}(2m)$ is the bosonic conformal Lie algebra with a basis $\{D,
P_{\dot\alpha}^{\ \beta}, K_{\alpha}^{\ \dot\beta}, M_{\alpha}^{\ \beta},
M_{\dot\alpha}^{\ \dot\beta}\}$. The other two summands in $\mathfrak{g}_{(0)}$ form the
internal symmetry algebra and their basis elements are denoted by $R_{I}^{\ J}$ and $R$,
respectively. Here, the indices $I,J = 1,...,\mathcal{N}$ are that of the fundamental
representation of $\mathfrak{sl}(\mathcal{N})$. Indices $\alpha,\dot\alpha=1,...,m$ are
that of the fundamental and the anti-fundamental representation of the rotation Lie algebra
inside $\mathfrak{sl}(2m)$. Of course, in the case $m=1$, there are no rotations and
these indices run over a one element set. Finally, the spaces $\mathfrak{q}_\pm$ and
$\mathfrak{s}_\pm$ have dimension $m\mathcal{N}$. The representations of $\mathfrak{k}$
which they carry are indicated by the type of indices of their generators
\begin{equation}
    \mathfrak{q}_+  = \text{span}\{Q_{\dot\alpha}^{\ J}\},\quad \mathfrak{q}_-  =
    \text{span}\{Q_I^{\ \beta}\},\quad \mathfrak{s}_+  = \text{span}\{S_{\alpha}^{\ J}\},
    \quad \mathfrak{s}_-  = \text{span}\{S_I^{\ \dot\beta}\}\ .
\end{equation}
The dual basis of $\mathfrak{g}_{(1)}^\ast$ to this one will be denoted by
$\{q^{\dot\alpha}_{\ J},q^I_{\ \beta},s^{\alpha}_{\ J},s^I_{\ \dot\beta}\}$.
Spaces $\mathfrak{g}_\pm$ carry representations of $\mathfrak{g}_{(0)}$ which are dual
to each other. Explicitly, the dual bases are
\begin{equation}
     (S_\alpha^{\ I})^* = Q_I^{\ \alpha},\quad (Q_{\dot\alpha}^{\ I})^* = S_I^{\ \dot\alpha}.
\end{equation}
The Lie superalgebra $\mathfrak{g}$ has a fundamental $2m+\mathcal{N}$-dimensional representation. We will denote by $E_i^{\ j}$ the matrix with 1 at position $(i,j)$
and zeros elsewhere, $i,j=1,...,2m+\mathcal{N}$. Such indices are split in three pieces $\dot\alpha,\alpha,I$, that is, we write
\begin{equation}
    A = A^i_{\ j} E_{i}^{\ j} = \begin{pmatrix}
    A^{\dot\alpha}_{\ \dot\beta} & A^{\dot\alpha}_{\ \beta} & A^{\dot\alpha}_{\ J}\\
     A^{\alpha}_{\ \dot\beta} & A^{\alpha}_{\ \beta} & A^{\alpha}_{\ J}\\
     (-1)^{|A^I_{\ \dot\beta}|}A^I_{\ \dot\beta} & (-1)^{|A^I_{\ \beta}|}A^I_{\ \beta} & (-1)^{|A^I_{\ J}|}A^I_{\ J}
    \end{pmatrix}.
\end{equation}
We can choose the generators so that the $\mathfrak{sl}(2m)$ and $\mathfrak{sl}(\mathcal{N})$ algebras sit in the top left and bottom right corners, respectively, while the subspaces
$\mathfrak{g}_\pm$ occupy the top right and bottom left corners. Schematically
\begin{equation}
    \begin{pmatrix}
    \mathfrak{k}\cap\mathfrak{sl}(2m) & \mathfrak{g}_1 & \mathfrak{q}_+ \\
    \mathfrak{g}_{-1} & \mathfrak{k}\cap\mathfrak{sl}(2m) & \mathfrak{s}_+ \\
    \mathfrak{s}_- & \mathfrak{q}_- & \mathfrak{sl}(\mathcal{N})
    \end{pmatrix}\ .
\end{equation}
For the precise definition of the fundamental representation, see the appendix B.\footnote{The early works \cite{Park:1997bq,Osborn:1998qu,Park:1999pd} use the same representation both of the Lie superalgebra and the supergroup and have inspired some of our calculations.}
\medskip

In the remainder of this subsection, we will derive expression for $y(x),z(x),t(x)$
appearing in eq.\ \eqref{matrix-identity} . In order to do this, we spell out the supermatrices
representing various factors in this equation. The Weyl inversion and its inverse take the form
\begin{equation}
   w = \begin{pmatrix}
    0 & -w^{\dot\alpha}_{\ \beta} & 0 \\
    -w^{\alpha}_{\ \dot\beta} & 0 & 0 \\
    0 & 0 & \delta^I_{\ J}
    \end{pmatrix},\quad  w^{-1} = \begin{pmatrix}
    0 & w^{\dot\alpha}_{\ \beta} & 0 \\
    w^{\alpha}_{\ \dot\beta} & 0 & 0 \\
    0 & 0 & \delta^I_{\ J}
    \end{pmatrix},
\end{equation}
where $-w^{\dot\alpha}_{\ \beta} = w^{\alpha}_{\ \dot\beta} = \sigma_2 $ for $m=2$ and $w^{\dot\alpha}_{\ \beta} = - w^{\alpha}_{\ \dot\beta}=1$ for $m=1$.

The superspace $\mathcal{M} = G/P$ is generated by variables $x^{\dot\alpha}_{\ \beta}, \theta^{\dot\alpha}_{\ J}, \bar\theta^{I}_{\ \beta}$, obeying the usual (anti)commutation relations. We see that
\begin{equation}
    m(x) = e^{x^{\dot\alpha}_{\ \beta}P_{\dot\alpha}^{\ \beta} + \theta^{\dot\alpha}_{\ J}Q_{\dot\alpha}^{\ J} + \bar\theta^{I}_{\ \beta} Q_I^{\ \beta}}=\begin{pmatrix}
    \delta^{\dot\alpha}_{\ \dot\beta} & X^{\dot\alpha}_{\ \beta} & \theta^{\dot\alpha}_{\ J}\\
    0 & \delta^{\alpha}_{\ \beta} & 0\\
    0 & -\bar\theta^{I}_{\ \beta} & \delta^{I}_{\ J}
    \end{pmatrix},\ \ \text{with} \ \ X^{\dot\alpha}_{\ \beta} = x^{\dot\alpha}_{\ \beta}-\frac12\theta^{\dot\alpha}_{\ K} \bar\theta^{K}_{\ \beta}\ .
\end{equation}
Using $w^{\dot\alpha}_{\ \delta} w^{\delta}_{\ \dot\beta} = -\delta^{\dot\alpha}_{\ \dot\beta}$ and $w^{\alpha}_{\ \dot\delta} w^{\dot\delta}_{\ \beta} = -\delta^{\alpha}_{\ \beta}$ we get for elements $n(x)$
\begin{equation}\label{elements-n}
    n(x) = w^{-1} m(x) w = \begin{pmatrix}
    \delta^{\dot\alpha}_{\ \dot\beta} & 0 & 0 \\
    -w^\alpha_{\ \dot\gamma}X^{\dot\gamma}_{\ \delta}w^{\delta}_{\ \dot\beta} & \delta^{\alpha}_{\ \beta} & w^{\alpha}_{\ \dot\gamma}\theta^{\dot\gamma}_{\ J}\\
    \bar\theta^I_{\ \delta} w^{\delta}_{\ \dot\beta} & 0 & \delta^{I}_{\ J}
    \end{pmatrix}.
\end{equation}
Finally, elements of the subgroup $K$ assume the form
\begin{equation}
    k(t) = \begin{pmatrix}
    e^{\frac{\mathcal{N}\kappa}{\mathcal{N}-2m}+\frac12\lambda}(r_1)^{\dot\alpha}_{\ \dot\beta} & 0 & 0 \\
    0 & e^{\frac{\mathcal{N}\kappa}{\mathcal{N}-2m}-\frac12\lambda}(r_2)^{\alpha}_{\ \beta} & 0 \\
    0 & 0 & e^{\frac{2m\kappa}{\mathcal{N}-2m}}U^{I}_{\ J}
    \end{pmatrix} \equiv \text{diag}(k_1,k_2,k_3)\ .
\end{equation}
Matrices $r_{1,2}$ are purely rotational. That is, they belong to $SL(2,\mathbb{C})$ for $m=2$ and are equal to 1 if $m=1$.

In the following we will suppress indices where no confusion can arise. They can be put back at any point by looking at what type of indices a certain object carries and contracting over the appropriate number and type of dummy indices. We shall agree to write $J=w^{\alpha}_{\ \dot\beta}$, then $-J^{-1}=w^{\dot\alpha}_{\ \beta}$. With these conventions, the above expressions can be rewritten as
\begin{equation}
    w= \begin{pmatrix}
    0 & J^{-1} & 0 \\
    -J & 0 & 0\\
    0 & 0 & 1
    \end{pmatrix},\ w^{-1} = \begin{pmatrix}
    0 & -J^{-1} & 0 \\
    J & 0 & 0\\
    0 & 0 & 1
    \end{pmatrix},\ m(x) = \begin{pmatrix}
    1 & X & \theta \\
    0 & 1 & 0\\
    0 & -\bar\theta & 1
    \end{pmatrix},\ n(x) = \begin{pmatrix}
    1 & 0 & 0 \\
    -JXJ & 1 & J\theta\\
    \bar\theta J & 0 & 1
    \end{pmatrix}\ .
\end{equation}
Therefore, the equation $w m(x) = m(y)n(z)k(t)$ reads
\begin{equation}
    \begin{pmatrix}
    0 & J^{-1} & 0 \\
    -J & -J X & -J\theta\\
    0 & -\bar\theta & 1
    \end{pmatrix} =  \begin{pmatrix}
    (1 - YJZJ + \eta\bar\zeta J)k_1 & Y k_2 & (YJ\zeta+ \eta) k_3\\
    -JZJ k_1 & k_2 & J\zeta k_3\\
    (\bar\eta JZJ + \bar\zeta J)k_1 & -\bar\eta k_2 & (1-\bar\eta J\zeta) k_3
    \end{pmatrix}\ .
\end{equation}
Here, the notation is $y = (y^{\dot\alpha}_{\ \beta},\eta^{\dot\alpha}_{\ J},\bar\eta^I_{\ \beta})$, $z = (z^{\dot\alpha}_{\ \beta},\zeta^{\dot\alpha}_{\ J},\bar\zeta^I_{\ \beta})$ and $Y,Z$ are introduced analogously to $X$. To write down the solution for $y,z$ and $t$ we introduce
\begin{equation}
    T = 1 + \bar\theta X^{-1}\theta,\quad \Lambda = 1+X^{-1}\theta\bar\theta\ .
\end{equation}
Then one observes that $\Lambda^{-1} = 1-X^{-1}\theta T^{-1}\bar\theta$. Using this, the solution to the above system is found
\begin{align}
    &(Y,\eta,\bar\eta) = (- (JXJ)^{-1},-(XJ)^{-1}\theta T^{-1},-\bar\theta(JX)^{-1}),\\
    &(Z,\zeta,\bar\zeta) = (- X - \theta\bar\theta,-\theta T^{-1},-\bar\theta\Lambda),\\
    &(k_1,k_2,k_3) = (-((X+\theta\bar\theta)J)^{-1},-JX,T)\ .
\end{align}
In particular $z^{\dot\alpha}_{\ \beta} = -x^{\dot\alpha}_{\ \beta}$, as in the bosonic theory.
This completes our analysis of the equation \eqref{matrix-identity}.

\subsection{Cartan decomposition and crossing factor}

Having determined the decomposition \eqref{matrix-identity}, we can turn to the crossing equations. We shall consider $s$- and $t$-channels and use superconformal transformations to set
\begin{equation}
    x_1 = (a e_1 + b e_2,\theta_1,\bar\theta_1),\quad x_2 = (0,0,0),\quad x_3 = (e_1,\theta_3,\bar\theta_3),\quad x_4 = (\infty e_1,0,0)\ .
\end{equation}
To write the crossing symmetry equations, one should consider the primed Cartan decomposition of $G$. We start from its unprimed cousin
\begin{equation}\label{param}
g = e^{q^I_{\ \beta} Q_I^{\ \beta} + s^I_{\ \dot\beta} S_{I}^{\ \dot\beta}} \, k_l  a(u_1,u_2)  k_r\, e^{q^{\dot\alpha}_{\ J}Q_{\dot\alpha}^{\ J} + s^{\alpha}_{\ J} S_{\alpha}^{\ J}}\ .
\end{equation}
In the fundamental representation this reads
\begin{equation}
    g = \begin{pmatrix}
    \delta^{\dot\alpha}_{\ \dot\gamma} & 0 & 0\\
    0 & \delta^{\alpha}_{\ \gamma} & 0\\
    -s^I_{\ \dot\gamma} & -q^I_{\ \gamma} & \delta^{I}_{\ K}
    \end{pmatrix}    \begin{pmatrix}
    e^{\frac{\mathcal{N}\kappa}{\mathcal{N}-2m}} (g_b)^{\dot\gamma}_{\ \dot\delta} & e^{\frac{\mathcal{N}\kappa}{\mathcal{N}-2m}} (g_b)^{\dot\gamma}_{\ \delta} & 0 \\
    e^{\frac{\mathcal{N}\kappa}{\mathcal{N}-2m}} (g_b)^{\gamma}_{\ \dot\delta} & e^{\frac{\mathcal{N}\kappa}{\mathcal{N}-2m}} (g_b)^{\gamma}_{\ \delta} & 0\\
    0 & 0 & e^{\frac{2m\kappa}{\mathcal{N}-2m}} U^K_{\ L}
    \end{pmatrix}     \begin{pmatrix}
    \delta^{\dot\delta}_{\ \dot\beta} & 0 & q^{\dot\delta}_{\ J} \\
    0 & \delta^{\delta}_{\ \beta} & s^{\delta}_{\ J} \\
    0 & 0 & \delta^{L}_{\ J}
    \end{pmatrix},
\end{equation}
where $g_b = k_l^b a k_r^b$ is an element of the bosonic conformal group. We see that the top left $2m\times 2m$ corner is simply a scalar multiple of $g_b$. When written without indices, generators of $\mathfrak{g}_-^\ast$ will carry a bar, to be distinguished from generators of $\mathfrak{g}_+^\ast$. With this convention, the above Cartan decomposition reads
\begin{equation}
    g = e^{\frac{\mathcal{N}\kappa}{\mathcal{N}-2m}}\begin{pmatrix}
   A & B & Aq+Bs\\
   C & D & Cq+Ds\\
    -\bar s A  -\bar q C & -\bar s B - \bar q D & e^{-\kappa} U-(\bar s A +\bar q C)q -(\bar s B + \bar q D)s
    \end{pmatrix}\ .
\end{equation}
Here $A,B,C,D$ are $m\times m$ blocks of $g_b$ and how to extract Cartan coordinates from them was explained in \cite{Buric:2019dfk}. The elements that we want to decompose are
\begin{equation}
    g_s(x_i) = n(y_{21})^{-1} m(x_{31}) n(y_{43}),\quad g_t(x_i) = n(y_{41})^{-1} m(x_{31}) n(y_{23})\ .
\end{equation}
As can be seen from the solutions of eq.\ \eqref{matrix-identity} found in the previous subsection, when $x$ is sent to $(\infty e_1,\theta,\bar\theta)$ then $y(x)=0$ and consequently $n(y(x))=1$. Therefore, in the special configuration that we chose, one has
\begin{equation}\label{gst-gauge-fixed}
    g_s(x_i) = n(y_{21})^{-1} m(x_{31}),\quad g_t(x_i) = m(x_{31}) n(y_{23})\ .
\end{equation}
Thus we are led to consider the decomposition of elements that take the general form $n(y)m(x)$ and $m(x)n(y')$. We treat these in turn. In the notation of the previous subsection
\begin{equation}
    n(y)m(x) =  \begin{pmatrix}
    1 & X & \theta\\
    - JYJ & 1 - JYJX - J\eta\bar\theta & J\eta - JYJ\theta\\
    \bar\eta J & \bar\eta JX - \bar\theta & 1 + \bar\eta J \theta
    \end{pmatrix}\ .
\end{equation}
One immediately finds
\begin{equation}\label{eqns1}
    s_s = (1 - J \eta \bar\theta)^{-1} J \eta,\ q_s = \theta - X s_s,\ \bar q_s = \bar\theta (1-J\eta\bar\theta)^{-1},\ \bar s_s = (\bar q_s J Y - \bar\eta)J,\ e^{\frac{2m\kappa_s}{\mathcal{N}-2m}}U_s = 1 + \bar\theta s_s\ .
\end{equation}
The last expression can be simplified by substituting for $s$ and performing the following manipulation
\begin{equation}
    1 + \bar\theta s_s = 1+ \bar\theta(1 - J \eta \bar\theta)^{-1} J \eta = 1 + \bar\theta\Big(\sum_{n=0}^{\infty} (J \eta\bar\theta)^n\Big)J\eta = \sum_{n=0}^{\infty}(\bar\theta J\eta)^n = (1-\bar\theta J\eta)^{-1}\ . \nonumber
\end{equation}
Therefore, taking the determinant of the last equation in \eqref{eqns1} gives
\begin{equation}
    e^{\frac{2m\mathcal{N}\kappa_s}{\mathcal{N}-2m}} = \text{det}(1 - \bar\theta J\eta)^{-1}\ .
\end{equation}
Next, by looking at determinants of top left four $m\times m$ blocks we obtain the coordinates associated with dilations
\begin{equation}
    e^{2(\lambda_{s,l} + \lambda_{s,r})} = -\text{det}(J^{-1}-YJX-\eta\bar\theta)^{-1},\quad e^{2(\lambda_{s,l} - \lambda_{s,r})} = \text{det}X\text{det}Y^{-1},
\end{equation}
as well as the coordinates $(u_1,u_2)$ of the abelian torus
\begin{align}
    & \sinh^2\frac{u^s_1}{2}\sinh^2\frac{u^s_2}{2} = \text{det}X\text{det}Y\text{det}(1 - \bar\theta J\eta),\\
    & \cosh^2\frac{u^s_1}{2} \cosh^2\frac{u^s_2}{2} = -\text{det}(J^{-1}-YJX-\eta\bar\theta)\text{det}(1 - \bar\theta J\eta)\ .
\end{align}
We have already put a label $s$ on the coordinates, as they are indeed the $s$-channel coordinates for appropriate choices of $x$ and $y$ as in eq.\ \eqref{gst-gauge-fixed}. For the other channel, we decompose
\begin{equation}
    m(x)n(y') =  \begin{pmatrix}
    1 - XJY'J +\theta\bar\eta' J  & X & XJ\eta'+\theta\\
    -JY'J & 1 & J\eta'\\
    \bar\theta JY'J+\bar\eta' J & -\bar\theta & 1-\bar\theta J \eta'
    \end{pmatrix}\ .
\end{equation}
Following similar steps as above, we find
\begin{equation}
    q_t = (1+\theta\bar\eta' J)^{-1}\theta,\ s_t=J(\eta'+Y'J q_t),\ \bar s_t = -\bar\eta' J (1+\theta\bar\eta' J)^{-1},\ \bar q_t = \bar\theta - \bar s_t X,\ e^{\frac{2m\kappa_t}{\mathcal{N}-2m}}U_t = 1 - \bar\eta J q_t,
\end{equation}
and therefore
\begin{equation}
    e^{\frac{2m\kappa_t}{\mathcal{N}-2m}} = \text{det}(1 + \bar\eta' J\theta)^{-1}\ .
\end{equation}
Dilation coordinates are now
\begin{equation}
    e^{2(\lambda_{t,l} + \lambda_{t,r})} = -\text{det}(J^{-1}-XJY'+\theta\bar\eta'),\quad e^{2(\lambda_{t,l} - \lambda_{t,r})} = \text{det}X\text{det}Y'^{-1}\ .
\end{equation}
Finally the coordinates on the abelian torus read
\begin{align}
    & \sinh^2\frac{u^t_1}{2}\sinh^2\frac{u^t_2}{2} = \text{det}X\text{det}Y' \text{det}(1+\bar\eta' J\theta),\\
    & \cosh^2\frac{u^t_1}{2} \cosh^2\frac{u^t_2}{2} = -\text{det}(J^{-1}-XJY'+\theta\bar\eta')\text{det}(1+\bar\eta' J\theta)\ .
\end{align}
Expressions written so far are sufficient to determine the crossing factor for fields which transform trivially under rotations and $SU(\mathcal{N})$ internal symmetries. For applications that we have in mind in this work these conditions are satisfied. Assuming that a field transforms trivially both under spatial rotations and $SU(\mathcal{N})$ internal symmetries, it is associated with a 1-dimensional representation $\rho_{\Delta,r}$ of $K$. Here $\Delta$ is the conformal weight and $r$ the $U(1)_R$-charge of the field. Our parametrisation of $K$ is such that
\begin{equation}
    \rho_{\Delta,r}(e^{\lambda D + \kappa R}e^{r^\alpha_{\ \beta} M_{\alpha}^{\ \beta} + r^{\dot\alpha}_{\ \dot\beta} M_{\dot\alpha}^{\ \dot\beta} + u^I_{\ J} R_I^{\ J}}) = e^{-\Delta\lambda + r\kappa}.
\end{equation}
Therefore, the tensor factors appearing in $\mathcal{M}_{st}$ are
\begin{align}
    & \rho_3(\kappa_3) = e^{\Delta_3(\lambda_{s,r}-\lambda_{t,r})},\ \ \rho_4(\kappa_4) = e^{-\Delta_4(\lambda_{t,l}+\lambda_{s,r})-r_4\kappa_t},\\[2mm]
    & \rho_1(\kappa_1) = e^{\Delta_1(\lambda_{t,l}-\lambda_{s,l})+r_1(\kappa_s-\kappa_t)},\ \ \rho_2(\kappa_2) = e^{\Delta_2(\lambda_{t,r}+\lambda_{s,l})+r_2\kappa_s}\rho_2(k(t_{23})k(t_{21})^{-1})\ .
\end{align}
In the last expression we have used that the middle two factors in $\kappa_4$ cancel out in our gauge. All the coordinates appearing on right hand sides of previous equations have been spelled out and one simply substitutes for them to find the product.

\subsection{Application to 4-dimensional $\mathcal{N}=1$ SCFTs}

Let us apply the results from previous two subsections to the complexified $\mathcal{N}=1$ superconformal algebra in $d=4$ dimensions, $\mathfrak{g} = \mathfrak{sl}(4|1)$. We use the same notation as above, only $\mathfrak{sl}(\mathcal{N})$ indices become redundant, as this summand disappears for $\mathcal{N}=1$.

The correlation function we want to consider is that of two long multiplets $\mathcal{O}$, along with one anti-chiral field $\bar\varphi_1$ and one chiral $ \varphi_3$, see eq.\ \eqref{4-point-function}. The fields have conformal weights $\Delta_i$ and $R$-charges $r_i$, and we assume that $\sum r_i = 0$. Therefore, we can write $r = r_1 + r_2 = - r_3 - r_4$. Chirality conditions further imply
\begin{equation}
    \Delta_1 = - \frac32 r_1, \quad \Delta_3 = \frac32 r_3\ .
\end{equation}
The general solution for $y(x)$ specialises in the case $m=2,\ \mathcal{N}=1$ to
\begin{align}
     y = - \frac{1+\frac{\Omega}{4\text{det}x}}{\text{det}x}x^{t},\ \eta = \frac{-i}{\text{det}x}\begin{pmatrix}
    x^{\dot2}_{\ 1}\theta^{\dot1} - x^{\dot1}_{\ 1}\theta^{\dot2} + \frac12 \bar\theta_1 \theta^{\dot 1}\theta^{\dot 2}\\
    x^{\dot2}_{\ 2}\theta^{\dot1} - x^{\dot1}_{\ 2}\theta^{\dot2} + \frac12 \bar\theta_2 \theta^{\dot 1}\theta^{\dot 2}
    \end{pmatrix},\ \bar\eta^T = \frac{-i}{\text{det}x}\begin{pmatrix}
    x^{\dot1}_{\ 1}\bar\theta_2 - x^{\dot1}_{\ 2}\bar\theta_1 - \frac12 \theta^{\dot1} \bar\theta_1\bar\theta_2\\
    x^{\dot2}_{\ 1}\bar\theta_2 - x^{\dot2}_{\ 2}\bar\theta_1 - \frac12 \theta^{\dot2} \bar\theta_1\bar\theta_2
    \end{pmatrix}\ .\nonumber
\end{align}
In these formulas, $x$ and $y$ denote $2\times2$ matrices of {\it bosonic} coordinates of super-points $x$ and $y$. This is a slight abuse of notation, but in any equation the meaning of symbols $x,y$ is clear from the context. By $\Omega$ we denote the element $\theta^{\dot1}\theta^{\dot2}\bar\theta_1\bar\theta_2$. The covariant derivatives, realising the right-regular action of $\mathfrak{m}$, read in our coordinates
\begin{equation}
    D_{\dot\alpha}^{\ I} = \partial_{\dot\alpha}^{\ I} + \frac12\bar\theta^{I}_{\ \beta}\partial_{\dot\alpha}^{\ \beta},\quad \bar D_{I}^{\ \alpha} = -\partial_{I}^{\ \alpha}-\frac12\theta^{\dot\beta}_{\ I}\partial_{\dot\beta}^{\ \alpha}\ .
\end{equation}
One can verify that they anti-commute with the right-invariant vector fields written in the appendix B. Let us introduce the corresponding chiral and anti-chiral coordinates
\begin{equation}
    x'^{\dot\alpha}_{\ \beta} = x^{\dot\alpha}_{\ \beta} + \frac12\theta^{\dot\alpha}_{\ I}\bar\theta^{I}_{\ \beta},\quad x''^{\dot\alpha}_{\ \beta} = x^{\dot\alpha}_{\ \beta} - \frac12\theta^{\dot\alpha}_{\ I}\bar\theta^{I}_{\ \beta}\ .
\end{equation}
We further set $\theta' = \theta'' = \theta$ and $\bar\theta' = \bar\theta'' = \bar\theta$. Then the following equalities hold
\begin{equation}
    D x'' = D\bar\theta'' = 0, \quad \bar D x' = \bar D \theta' = 0\ .
\end{equation}
The chirality conditions satisfied by the fields allow us to set $\theta_1$ and $\bar\theta_3$ to zero. Let us write $\alpha = a + ib$, $\alpha^\ast = a - ib$ and fix the insertion points to positions as explained in the previous subsection. Further, we write $y = - y_{21}$ and $y'=y_{23}$. Then a computation gives
\begin{align}
    & y = \Big( \begin{pmatrix}
    -1/\alpha^\ast & 0\\
    0 & 1/\alpha
    \end{pmatrix},\ \begin{pmatrix}
    0\\
    0
    \end{pmatrix},\ i\begin{pmatrix}
    (\bar\theta_1)_2/\alpha^\ast\\
    (\bar\theta_1)_1/\alpha
    \end{pmatrix}^t\Big),\ y' = \Big( \begin{pmatrix}
    1 & 0\\
     0 & -1
    \end{pmatrix},\ i\begin{pmatrix}
    (\theta_3)^{\dot2}\\
    (\theta_3)^{\dot1}
    \end{pmatrix},\ \begin{pmatrix}
    0\\
    0
    \end{pmatrix}^t\Big)\ .
\end{align}
Next, the factor $m(x_{31})$ is found
\begin{equation}
    m(x_{31}) = \begin{pmatrix}
    1 & X & \theta_3\\
    0 & 1 & 0\\
    0 & \bar\theta_1 & 1
    \end{pmatrix}, \ \text{with}\ X = X_3 - X_1 = \begin{pmatrix}
    -1 + \alpha & 0\\
    0 & 1 - \alpha^\ast
    \end{pmatrix}\ .
\end{equation}
We are now ready to consider the Cartan decomposition of $n(y)m(x)$ and $m(x)n(y')$. Using the formulas of the previous subsection, the fermionic coordinates and dilation coordinates are
\begin{align}
    & q_s = q_t = \theta_3,\quad \bar q_s = \bar q_t = -\bar\theta_1,\quad s_s = \bar s_s = s_t = \bar s_t = 0,\\
    & e^{4\lambda_{s,l}} = \alpha^2(\alpha^\ast)^2(1-\alpha)(1-\alpha^\ast),\quad e^{4\lambda_{s,r}} = \frac{1}{(1-\alpha)(1-\alpha^\ast)},\\
    & e^{4\lambda_{t,l}} = \alpha\alpha^\ast(1-\alpha)(1-\alpha^\ast),\quad e^{4\lambda_{t,r}} = \frac{\alpha\alpha^\ast}{(1-\alpha)(1-\alpha^\ast)}\ .
\end{align}
The other factors that appear in $k_{s/t,l/r}$, which are products of rotations and $R$-symmetry transformations, assume the following diagonal form
\begin{equation}
    r_{s/t,l} = \begin{pmatrix}
    L_{s/t} & 0 & 0\\
    0 & L_{s/t}^{-1} & 0\\
    0 & 0 & 1
    \end{pmatrix},\ r_{s/t,r}=\begin{pmatrix}
    R_{s/t} & 0 & 0\\
    0 & R_{s/t}^{-1} & 0\\
    0 & 0 & 1
    \end{pmatrix}, \ \ \text{with} \ \ L_{s,t} = \begin{pmatrix}
    l_{s,t} & 0\\
    0 & l_{s,t}^{-1}
    \end{pmatrix},\ \ R_{s,t} = \begin{pmatrix}
    r_{s,t} & 0\\
    0 & r_{s,t}^{-1}
    \end{pmatrix},
\end{equation}
and $l_{s,t},\ r_{s,t}$ are in turn given by
\begin{equation}
    l_s = \Big(\frac{\alpha^2(1-\alpha)}{(\alpha^\ast)^2(1-\alpha^\ast)}\Big)^{1/8},\ r_s = \Big(\frac{1-\alpha^\ast}{1-\alpha}\Big)^{1/8},\ l_t = \sqrt{-i}\Big(\frac{\alpha(1-\alpha)}{\alpha^\ast(1-\alpha^\ast)}\Big)^{1/8},\ r_t = \frac{1}{\sqrt{-i}}\Big(\frac{\alpha(1-\alpha^\ast)}{\alpha^\ast(1-\alpha)}\Big)^{1/8}\ . \nonumber
\end{equation}
Finally, the coordinates on the torus are
\begin{equation}
    \cosh^2\frac{u_1^s}{2} = \frac{1}{\alpha},\quad \cosh^2\frac{u_2^s}{2} = \frac{1}{\alpha^\ast},\quad \cosh^2\frac{u_1^t}{2} = \alpha,\quad \cosh^2\frac{u_2^t}{2} = \alpha^\ast\ .
\end{equation}
This completes the determination of Cartan coordinates of the elements $g_s$ and $g_t$. To find the matrix $\mathcal{M}_{st}$, it is still required to determine $k(t_{21})$ and $k(t_{23})$. These are
\begin{equation}
    k(t_{21}) = (\alpha\alpha^\ast)^{-D},\quad k(t_{23}) = 1\ .
\end{equation}
This allows for the computation of factors appearing in $\mathcal{M}_{st}$. The computation gives
\begin{align}
    \rho_i(\kappa_i) = (\alpha\alpha^\ast)^{-\frac{\Delta_i}{4}}\ .
\end{align}
To derive the crossing equations, there is one remaining step, namely to perform the expansion in nilpotent invariants in both channels. In order to do this, we need to switch to the primed Cartan coordinates by moving the exponentials containing fermionic variables past the elements of the left and right $K$-subgroups. We have in both channels that $s' = \bar s' = 0$ and
\begin{align}
     \bar q_s' = -\bar\theta_1 L_s^{-1} e^{-\frac12\lambda_{s,l}},\quad q_s' = R_s e^{\frac12\lambda_{s,r}}\theta_3,\quad \bar q'_t = -\bar\theta_1 L_t^{-1} e^{-\frac12\lambda_{t,l}},\quad q_t' = R_t e^{\frac12\lambda_{t,r}}\theta_3\ .
\end{align}
Recall that $B$ is the commutant in $G_{(0)}$ of the 2-dimensional abelian group $A$. In the case at hand, $B=SO(2)\times SO(2)$ and Lie algebras of $A$ and $B$ are
\begin{equation}
    \mathfrak{a} = \text{span}\{P_1 + K_1,\ P_2 - K_2\},\quad \mathfrak{b} = \text{span}\{R,M_1^{\ 1} + M_{\dot1}^{\ \dot1}\}\ .
\end{equation}
Irreducible finite-dimensional representations of $\mathfrak{k}$ are labelled by two spins, a conformal weight and an $R$-charge, $(j_1,j_2)^{\Delta}_r$. In such notation, the four modules $\mathfrak{q}_\pm,\mathfrak{s}_\pm$ are
\begin{equation}
    \mathfrak{q}_+ = (0,1/2)^{1/2}_1,\quad \mathfrak{q}_- = (1/2,0)^{1/2}_{-1},\quad \mathfrak{s}_+ = (1/2,0)^{-1/2}_1,\quad \mathfrak{s}_- = (0,1/2)^{-1/2}_{-1}\ .
\end{equation}
According to our general theory, \cite{Buric:2019rms}, blocks for the correlation function \eqref{4-point-function} are functions on the double coset with values in the space
\begin{equation}
\left( V_{(12)} \otimes \Lambda \mathfrak{q}_- \otimes V_{(34)} \otimes \Lambda \mathfrak{q}_+ \right)^{\mathfrak{b}} = \left( \Lambda \mathfrak{q} \right)^\mathfrak{b}\ . \label{eq:d4space}
\end{equation}
Under the action of $\mathfrak{k}$ the 16-dimensional exterior algebra inside the brackets transforms as
\begin{equation}
    \Lambda \mathfrak{q} \cong \mathbb{1}^0_0 \oplus \mathbb{1}^1_2 \oplus \mathbb{1}^{1}_{-2} \oplus \mathbb{1}^2_0 \oplus (1/2,0)^{1/2}_{-1} \oplus (0,1/2)^{1/2}_1 \oplus (1/2,1/2)^1_0\oplus (1/2,0)^{3/2}_{1} \oplus (0,1/2)^{3/2}_{-1}\ . \nonumber
\end{equation}
We have written $\mathbb{1}$ for the trivial representation of $SU(2)\times SU(2)$. Two of the singlets are $\mathfrak{b}$-invariant and the 4-dimensional representation $(1/2,1/2)$ contains a 2-dimensional invariant subspace. Hence, the space of invariants is 4-dimensional and spanned by
\begin{equation}\label{B-invariants}
    W_{\bar\varphi\mathcal{O}\varphi\mathcal{O}} = (\Lambda\mathfrak{q})^B = \text{span}\{1,Q_{\dot 1}Q^1,Q_{\dot 2}Q^2, Q_{\dot 1}Q^1 Q_{\dot 2}Q^2\}\ .
\end{equation}
Indeed, from the bracket relations given in the appendix, one checks that these combinations of generators commute, in the universal enveloping algebra $U(\mathfrak{g})$, with $M_1^{\ 1} + M_{\dot1}^{\ \dot1}$ and $R$. In the two channels, the invariant combinations read
\begin{align}
    & (\bar q'_t)_1 (q'_t)^{\dot 1} = -i (1-\alpha)^{-1/2} (\bar\theta_1)_1 (\theta_3)^{\dot 1},\ (\bar q'_t)_2 (q'_s)^{\dot 2} = i (1-\alpha^\ast)^{-1/2} (\bar\theta_1)_2 (\theta_3)^{\dot 2},\\
    & (\bar q'_s)_1 (q'_s)^{\dot 1} = - \alpha^{-1/2} (1-\alpha)^{-1/2} (\bar\theta_1)_1 (\theta_3)^{\dot 1},\ (\bar q'_s)_2 (q'_s)^{\dot 2} = -(\alpha^\ast)^{-1/2} (1-\alpha^\ast)^{-1/2} (\bar\theta_1)_2 (\theta_3)^{\dot 2}\ .
\end{align}
Putting everything together, the crossing factor between $s$- and $t$-channels reads
\begin{equation} \label{eq:Mst}
    M_{st} = (\alpha\alpha^\ast)^{\frac74-\frac14 \sum\Delta_i}\begin{pmatrix}
    1 & 0 & 0 & 0\\
    0 & i\sqrt{\alpha} & 0 & 0\\
    0 & 0 & -i\sqrt{\alpha^\ast} & 0\\
    0 & 0 & 0 & \sqrt{\alpha\alpha^\ast}
    \end{pmatrix}^{-1}\ .
\end{equation}
The factor $(\alpha\alpha^\ast)^{7/4}$ is the ratio of Haar measure densities in the two channels. We may observe that variables $\alpha,\alpha^\ast$ are related to the usual variables $z,\bar z$ by
\begin{equation} \label{eq:alpha}
    \alpha = \frac{z}{z-1}, \quad \alpha^\ast = \frac{\bar z}{\bar z - 1}\ .
\end{equation}
Thus the top left entry of the crossing matrix is the one that we would get in the bosonic theory, \cite{Buric:2020buk}, as expected.

\section{Superconformal Blocks for 4-dimensional $\mathcal{N}=1$ SCFTs}

In this section we will compute the superconformal blocks for the correlator \eqref{4-point-function},
which will allow us to write the associated crossing equation. According to the general strategy of the
group theoretical approach, the relevant Casimir operators for conformal blocks descend from the Laplace
operator on the (super)conformal group. Hence, the first subsection is devoted to the construction of
the Laplacian in the (unprimed) Cartan coordinates on a supergroup of type I which were introduced in
section 2.3. To obtain the Casimir operators we need to restrict the full Laplacian from the set of
all (vector valued) functions on $G$ to the subspace of so-called $K$-spherical functions, see below.
The reduction of the Laplacian to the relevant $K$-spherical functions on $SL(4|1)$ is performed in
the second subsection. In our group theoretical approach, the reduced (Casimir) operators take the
form of a matrix Schr\" odinger operator $H$. Its eigenfunctions are constructed in the third
subsection in terms of ordinary (Gauss-type) hypergeometric functions. The precise relation of
these eigenfunctions with superconformal blocks for the various operators that contribute to the
correlation function \eqref{4-point-function} is detailed in the final subsection before we
derive the crossing symmetry equations we stated in the introduction.

\subsection{The Laplacian in Cartan coordinates}

In our harmonic analysis approach to superconformal correlations, four-point functions are
represented as so-called $K$-spherical functions on the superconformal group. These are
vector-valued functions covariant with respect to both left and right regular action of
the subgroup $K\subset G$, with covariance laws that are determined by representations
that label the four fields of the correlator, see \cite{Buric:2020buk}. When we perform
the map from correlation functions to $K$-spherical functions, the superconformal Casimir
operator that is conventionally used to characterize the superconformal blocks is carried
to the Laplace-Beltrami operator on $G$. Thus the question of computing partial waves
becomes one in harmonic analysis.

The Laplacian on a Lie group or a supergroup can be constructed as the quadratic
Casimir made out of right-invariant vector fields (or left-invariant, the two
prescriptions give the same operator). In any given coordinate system it is a
second order differential operator that is typically very complicated. However,
since it commutes with the left and right regular actions, the Laplacian acts
within the space of $K$-spherical functions and, since the work of Berezin
\cite{Berezin}, is known to reduce to a simple operator on this space. These
classical results apply to ordinary Lie groups, but as we have shown in
\cite{Buric:2019rms}, admit a very satisfactory extension to include supersymmetry
of type I. The simplicity of the Laplacian now follows from its expression in the
unprimed Cartan coordinates that we now review.

Let $\{X^A\}$ be a basis for a type I Lie superalgebra $\mathfrak{g}$. We write
$\{X^A\} = \{X^a,X^\mu,X_\mu\}$ with
\begin{equation}
    \mathfrak{g}_{(0)} = \text{span}\{X^a\}, \quad \mathfrak{g}_+ = \text{span}\{X^\mu\},
    \quad \mathfrak{g}_- = \text{span}\{X_\mu\}\ .
\end{equation}
Spaces $\mathfrak{g}_+$ and $\mathfrak{g}_-$ carry representations of the even
subgroup $G_{(0)}$ (under the adjoint action) that are dual to each other
\begin{equation}\label{adjoint-representation}
    g_{(0)} X^\mu g_{(0)}^{-1} = \pi(g_{(0)})^\mu_{\ \nu} X^\nu, \quad
    g_{(0)} X_\nu g_{(0)}^{-1} = \pi(g_{(0)}^{-1})^\mu_{\ \nu} X_\mu\ .
\end{equation}
Let us denote by $K^{ab}$ the Killing form of the even subalgebra $\mathfrak{g}_{(0)}$.
It can be used to construct the quadratic Casimir element
\begin{equation}
    C_2 = K_{ab} X^a X^b - X^\mu X_\mu + X_\mu X^\mu\ .
\end{equation}
In the present notation, the (unprimed) Cartan decomposition of $G$ may be written as
\begin{equation}
    g  = e^{x^\mu X_\mu} g_{(0)} e^{x_\nu X^\nu}\ .
\end{equation}
The middle factor $g_{(0)}$ is an element of the underlying Lie group $G_{(0)} =
G_{bos}\times U$. We use Cartan coordinates for $G_{bos}$ as in \cite{Buric:2019dfk}
and some arbitrary coordinates for the internal symmetry group $U$. Now a short
computation of right-invariant vector fields leads to the following expression
for the Laplacian, see e.g. \cite{Quella:2007hr},
\begin{equation}\label{Laplacian}
    \mathcal{R}_{C_2} =  \mathcal{R}_{C_2}^{(0)} -
    2\pi(g_{(0)}^{-1})^\mu_{\ \nu}\partial_{x_\nu}\partial_{x^\mu} - K_{ab}
    \pi(X^a)^\mu_{\ \mu}\mathcal{R}^{(0)}_{X^b}\ .
\end{equation}
This is a remarkable formula. The first term on the right-hand side is the Laplacian
of the underlying Lie group $G_{(0)}$. The second term involves second order derivatives
with respect to fermionic coordinates only, with coefficients depending on bosonic
coordinates. Hence, this term is nilpotent. In the last term, the first order differential
operators $\mathcal{R}^{(0)}_{X^b}$ denote right-invariant vector fields on $G_{(0)}$. These
are multiplied with the trace of the bosonic generators $X^a$ in the representation on
fermionic generators.

In what follows we will expand scalar functions on the supergroup $G$ in the fermionic coordinates,
thereby writing them as vector-valued functions on $G_{(0)}$. In the process, the Laplacian on the
supergroup turns into a matrix-valued operator which, and after restriction to the appropriate subspace
of $K$-spherical functions, can be reduced to the two-dimensional space $A$ that is parametrised by
$u_1$ and $u_2$. Once this is done, the bosonic Laplacian becomes a matrix-valued Calogero-Sutherland
Hamiltonian whose eigenfunctions are spinning bosonic conformal blocks, \cite{Schomerus:2016epl,
Schomerus:2017eny}. Parameters of the Hamiltonian depend both on the quantum numbers of fields in the
correlator and on the representation of $K$ carried by $\mathfrak{g}_+$, see \cite{Buric:2019rms}.
This follows from the fact that the unprimed fermionic Cartan coordinates transform non-trivially
under the regular action of $K$. As can be readily seen from eq.\ \eqref{Laplacian}, the second term
on the right-hand side reduces to a nilpotent, upper triangular matrix of functions in the two
variables $u_1$ and $u_2$. The third term can at most have one non-zero contribution, which arises
from the $U(1)_R$-generator, the only one that is not traceless in the representation $\pi$. Since
this generator is also used in the formulation of $K$-covariance laws, the whole term reduces to a
matrix of constants. These comments may become more transparent in the example of $SL(4|1)$ that
we will now treat in detail.

\subsection{Casimir equations for 4-dimensional $\mathcal{N}=1$ SCFTs}

Let us now apply the formula \eqref{Laplacian} for the Laplacian to derive the Casimir equations for
the correlation function \eqref{4-point-function}. We will make use of both primed and unprimed
Cartan coordinates on the supergroup $SL(4|1)$. The unprimed ones read
\begin{equation}\label{Cartan-coords-1}
    g = e^{q_\alpha Q^\alpha + s_{\dot\alpha} S^{\dot\alpha}} e^{\kappa R} e^{\lambda_l D} r_l
    e^{\frac{u_1+u_2}{4}(P_1+K_1) - i\frac{u_1-u_2}{4}(P_2-K_2)} r_r
    e^{\lambda_r D}e^{q^{\dot\alpha}Q_{\dot\alpha} + s^\alpha S_{\alpha}},
\end{equation}
with
\begin{equation}
r_l = e^{\varphi^l_1 X_1} e^{\theta^l_1 Z_1} e^{\psi^l_1 X_1} e^{\varphi^l_2 X_2} e^{\theta^l_2 Z_2}
e^{\psi^l_2 X_2}\ ,\quad  r_r = e^{\varphi^r_1 X_1} e^{\theta^r_1 Z_1} e^{\psi^r_1 X_1}
e^{\varphi^r_2 X_2} e^{\theta^r_2 Z_2} e^{\psi^r_2 X_2},
\end{equation}
and $\psi_2^l = -\psi_1^l$. These coordinates are useful because they make the Laplacian take a
particularly simple form. On the other hand, the primed Cartan coordinates
\begin{equation}\label{Cartan-coords-2}
    g = e^{\kappa R} e^{\lambda_l D} r_l e^{q'_\alpha Q^\alpha + s'_{\dot\alpha} S^{\dot\alpha}}
    e^{\frac{u_1+u_2}{4}(P_1+K_1) - i\frac{u_1-u_2}{4}(P_2-K_2)} e^{q'^{\dot\alpha}Q_{\dot\alpha}
    + s'^\alpha S_{\alpha}} r_r e^{\lambda_r D},
\end{equation}
are well suited to formulate the restriction to $K$-spherical functions, i.e. for the formulation
of the left and right $K$-covariance laws. The explicit relation between coordinate systems
\eqref{Cartan-coords-1} and \eqref{Cartan-coords-2} is written in the appendix C.

For the four-point function \eqref{4-point-function}, the associated covariant functions $f$ on
the supergroup satisfy covariance properties
\begin{equation}\label{covariance-conditions-1}
    (\partial_{\lambda'_l} - 2a) f = (\partial_{\lambda'_r} - 2b) f = (\partial_{\kappa'} - r) f
    = \partial_{\varphi_1^{'l}} f = ... = \partial_{\psi_2^{'r}} f = 0\ .
\end{equation}
The parameters $a$ and $b$ are related to conformal weights of the fields in the correlation function
by $2a = \Delta_2 - \Delta_1$ and $2b = \Delta_3 - \Delta_4$. As we explained above, $\lambda_l' =
\lambda_l$ and similarly for all other variables that appear in eq.\ \eqref{covariance-conditions-1},
but the partial derivatives of course depend on the full system of coordinates and $\partial_{\lambda'_l}
\neq\partial_{\lambda_l}$ etc.  Due to the shortening of operators at positions $1$ and $3$, there are
further differential equations that $f$ obeys, namely
\begin{equation}\label{covariance-conditions-2}
    \partial_{s_{\dot\alpha}} f = \partial_{s^\alpha} f = 0\ .
\end{equation}
Having described covariance properties of $f$, we can perform the reduction of the Laplacian. From now
on, we focus on one channel, say the $s$-channel, the discussion for the other one being entirely
analogous. Upon expansion in the Grassmann coordinates, the function $f$ is regarded as a vector-valued
function on the underlying Lie group
\begin{equation}
    f : G_{(0)} \xrightarrow{} \Lambda\mathfrak{q}\ .
\end{equation}
On the two-dimensional abelian group $A$ generated by $\{P_1+K_1,P_2-K_2\}$, the function $f$ restricts
to $\omega^{1/2} G$ with
\begin{equation}\label{F-on-the-torus}
    G = G^{(1)}(u_i) + G^{(2)}(u_i) q_1 q^{\dot1} + G^{(3)}(u_i) q_2 q^{\dot2} + G^{(4)}(u_i) q_1 q^{\dot1}
    q_2 q^{\dot2}\ .
\end{equation}
The conventional factor $\omega^{1/2}$ is defined in \cite{Buric:2019dfk}. Other components of $G$ vanish
due to requirements of $B$-invariance. After the reduction, the Laplacian restricts to the operator
\begin{equation}\label{Hamiltonian}
    H = H_0 + A,
\end{equation}
that acts on $G$ according to the general theory from \cite{Buric:2019rms}. Let us first apply this theory
to find the {\it unperturbed} part $H_0$. This is the $4\times4$ matrix of differential operators
\begin{equation}\label{H0}
    H_0 = -\begin{pmatrix}
H_{\sc}^{a,b} + \frac{3}{16} r^2 + \frac34 r & 0 & 0 \\
0 & H^{a+\frac14,b-\frac14}_{\frac12}+\frac{3}{16} (r-1)^2 + \frac34 (r-1) & 0 \\
0 & 0 & H_{\sc}^{a+\frac12,b-\frac12} + \frac{3}{16} (r-2)^2 + \frac34 (r-2)
\end{pmatrix}\ .
\end{equation}
Here the operators on the diagonal are the scalar and the seed Calogero-Sutherland Hamiltonians. Explicitly
\begin{equation} \label{eq:Hsc}
    H^{a,b}_{\sc}=-\frac{1}{2}\partial^2_{u_1} - \frac{1}{2}\partial^2_{u_2} + \frac{1}{2}\left(\frac{(a+b)^2-\frac{1}{4}}{\sinh^2 u_1}-\frac{a b}{\sinh^2\frac{u_1}{2}} +\frac{(a+b)^2-\frac{1}{4}}{\sinh^2 u_2}-\frac{a b}{\sinh^2\frac{u_2}{2}}\right)+\frac{5}{4},
\end{equation}
and
\begin{gather}\label{eq:Hseed}
  H^{a,b}_{\frac{1}{2}}=\begin{pmatrix}
H^{a,b}_{sc} -\frac{1}{16} & 0 \\
0 & H^{a,b}_{sc}-\frac{1}{16}
\end{pmatrix}+\\[2mm]
\frac{1}{32}\omega\begin{pmatrix}
\frac{1}{\sinh^2\frac{u_1}{2}}+\frac{1}{\sinh^2\frac{u_2}{2}}+\frac{4}{\sinh^2\frac{u_1-u_2}{4}}-
\frac{4}{\cosh^2\frac{u_1+u_2}{4}} & 4(b-a)\left(\frac{1}{\sinh^2\frac{u_1}{2}}-\frac{1}{\sinh^2\frac{u_2}{2}}\right) \\ \nonumber
4(b-a) \left( \frac{1}{\sinh^2\frac{u_1}{2}}-\frac{1}{\sinh^2\frac{u_2}{2}}\right) & \frac{1}{\sinh^2\frac{u_1}{2}}+
\frac{1}{\sinh^2\frac{u_2}{2}}+\frac{4}{\sinh^2\frac{u_1+u_2}{4}}-\frac{4}{\cosh^2\frac{u_1-u_2}{4}}
\end{pmatrix}\omega^{-1},
\end{gather}
with
\begin{equation}
    \omega = \frac{1}{\sqrt{2}}\begin{pmatrix}
    1 & -1\\
    1 & 1
    \end{pmatrix}\ .
\end{equation}
Before we spell out the form of the nilpotent term $A$, let us make a few comments on the derivation of
$H_0$. We focus on the most non-trivial part of $H_0$, namely the seed Hamiltonian. Discussion of the
two scalar Hamiltonians then follows by similar arguments. The covariant function $f$ contains among
its 16 components the 4-component function
\begin{equation}
    f_{-+} : G_{(0)} \xrightarrow{} \mathfrak{q}_- \otimes \mathfrak{q}_+, \quad f_{-+} =
    f^{\alpha}_{\ \dot\alpha} q_\alpha q^{\dot\alpha}\ .
\end{equation}
We can express $f_{-+}$ in the primed Cartan coordinates as
\begin{equation}\label{seed-covariance-law}
    f_{-+} = e^{-\kappa+\frac12\lambda_l-\frac12\lambda_r} f^{\alpha}_{\ \dot\alpha} \mathcal{L}_{\alpha}^{\ \beta}
    \mathcal{R}^{\dot\alpha}_{\ \dot\beta} q'_\beta q'^{\dot\beta}\ .
\end{equation}
The last equation defines $SU(2)$ matrices $\mathcal{L},\mathcal{R}$. Their explicit form is easily
written down using the formulas of the appendix C. In the primed Cartan coordinates, the dilation
covariance laws simply read
\begin{equation}
    f_i(e^{\kappa R + \lambda_l D}g_{(0)}e^{\lambda_r D}) = e^{r\kappa + 2a \lambda_l + 2b\lambda_r} f_i(g_{(0)})\ .
\end{equation}
We have performed the expansion in fermionic primed Cartan coordinates. The index $i$ denotes any
component of $f^{\alpha}_{\ \dot\alpha}$. Thus, from eq. \eqref{seed-covariance-law} we see that
$a$ and $b$ receive shifts by $\pm1/4$, respectively. The two constant terms involving the $r$-charge
in the upper left matrix element of $H_0$ come from $-R^2/4$ in the quadratic Casimir and from the
third term in the Laplacian that involves the trace of $R$ in the representation $\mathfrak{g}_+$,
respectively. Again, in the seed Hamiltonian, $r$ is shifted by $-1$ due to eq.\
\eqref{seed-covariance-law}. Next, consider the covariance law with respect to rotations.
Similarly as in \cite{Schomerus:2017eny}, these lead us to the seed Hamiltonian $H_{1/2}$
(matrices $\mathcal{L},\mathcal{R}$ here are different from those used in \cite{Schomerus:2017eny}
but so is the projector $\mathcal{P}$, and they lead to the same Hamiltonian. The two calculations
are related by a change of basis).
\smallskip

Having described the unperturbed Hamiltonian, let us turn to the perturbation $A$, which reads
\begin{equation}\label{nilpotent}
    A = -2\begin{pmatrix}
0& a^1_{\ 3} & a^2_{\ 4} & 0 \\
0 & 0 & 0  & a^2_{\ 4} \\
0 & 0 & 0 & a^1_{\ 3}\\
0 & 0 & 0 & 0
\end{pmatrix} = -2\begin{pmatrix}
0& \sinh\frac{u_1}{2} & -\sinh\frac{u_2}{2} & 0 \\
0 & 0 & 0  & -\sinh \frac{u_2}{2}  \\
0 & 0 & 0 & \sinh \frac{u_1}{2}\\
0 & 0 & 0 & 0
\end{pmatrix}.
\end{equation}
To derive this, notice that in the representation $\mathfrak{g}_+$ of $G_{(0)}$, with the basis
$\{X^\mu\}=\{Q_{\dot1},Q_{\dot2},S_1,S_2\}$ (and with this order of basis vectors) the element
$a(u_1,u_2)$ is given by
\begin{equation}
   \pi\left(a(u_1,u_2)^{-1}\right)^\mu_{\ \nu} = \begin{pmatrix}
\cosh\frac{u_1}{2}& 0 & \sinh\frac{u_1}{2} & 0\\
0 & \cosh\frac{u_2}{2} & 0  & -\sinh\frac{u_2}{2}\\
\sinh\frac{u_1}{2} & 0 & \cosh\frac{u_1}{2} & 0\\
0 & -\sinh\frac{u_2}{2} & 0 & \cosh\frac{u_2}{2}
\end{pmatrix} \equiv (a^{\mu}_{\ \nu})\ .
\end{equation}
Basis of $\mathfrak{g}_-$ that is dual to the above basis of $\mathfrak{g}_+$ is $\{S^{\dot1},
S^{\dot2},Q^1,Q^2\}$. Now, an inspection of the second term in eq.\ \eqref{Laplacian} leads
to eq.\ \eqref{nilpotent}.

To summarise, the Laplacian eigenvalue equation on $SL(4|1)$ reduces on the space of $K$-spherical
functions to the eigenvalue problem of the operator $H$ written in eq.\ \eqref{Hamiltonian}. The
two ingredients of this operator are a $4\times 4$ matrix of differential operators $H_0$ and a
nilpotent matrix of functions $A$ in the two cross rations, see eqs.\ \eqref{H0} and
\eqref{nilpotent}. We now turn to eigenfunctions of $H$, that is, to the construction of
the superconformal blocks for the correlator \eqref{4-point-function}.

\subsection{Construction of the superconformal blocks}
\def\c{c}
\def\tc{c}
\def\tgamma{\gamma}
\def\s{\sigma}

Having set up the Casimir equation for our superblocks in Calogero-Sutherland gauge we now come to
the main task of this section, namely to solve these equations and thereby to construct the associated
superconformal blocks. Our formulas will be entirely explicit in that we construct the superblocks
as a finite linear combination of a set of known special functions. These functions appear already
as (bosonic) conformal blocks of the component fields in the supermultiplets or, put differently,
in the solution of the eigenvalue problem for the unperturbed Hamiltonian $H_0$. To set the stage,
let us introduce the eigenfunctions for the two different differential operators that appear along
the diagonal of $H_0$, see eqs.\ \eqref{eq:Hsc} and \eqref{eq:Hseed},
\begin{align}\label{eigenvalues-1}
     H^{a,b}_{\sc}\, \phi^{a,b}_{\Delta,l} & = C_{\Delta,l}^{\sc}\, \phi^{a,b}_{\Delta,l},\ \ \
     \quad \quad \quad C_{\Delta,l}^{\sc}=-\frac14\Delta(\Delta-4)-\frac14l(l+2),\\
     H^{a,b}_{\frac{1}{2}}\, \Psi^{(a,b)}_{\pm,\Delta,l} & = C^{\seed}_{\Delta,l}\, \Psi^{(a,b)}_{\pm,\Delta,l},\
    \ \  \quad C^{\seed}_{\Delta,l} = -\frac14\Delta(\Delta-4)-\frac14 l(l+3)-\frac38\ . \label{eigenvalues-2}
\end{align}
The eigenfunctions $\phi^{a,b}_{\Delta,l}$ are related to the usual scalar conformal blocks through
a simple transformation, see \cite{Isachenkov:2017qgn}. The precise relation of the eigenfunctions
$\Psi^{a,b}_{\pm,\Delta,l}$ with the simplest spinning seed conformal blocks of \cite{Echeverri:2016dun}
was found in \cite{Schomerus:2017eny}. Recall that the eigenvalue equation for $\Psi$ is a matrix
equation and hence $\Psi$ has two components. The subscript $\pm$ on $\Psi$ labels two linearly
independent solutions with the same eigenvalue. Explicit formulas for these functions can be found
in the appendix E. They can all be written in terms of Gauss' hypergeometric function as finite
sums of products thereof. These concrete formulas are needed e.g. for numerical evaluation of the
crossing symmetry constraints we spell out below. But for our derivation of these constraints we
only need a few very basic formulas that tell us how they behave under multiplication with the
matrix elements of the nilpotent potential $A$ that was given in eq.\ \eqref{nilpotent},
\begin{align} \label{eq:aux1}
     &2\begin{pmatrix}
    -\sinh\frac{u_1}{2}\  , \ & \sinh\frac{u_2}{2}
    \end{pmatrix}\Psi^{a+\frac14,b-\frac14}_{+,\Delta+1,l} =  \gamma^{+1}_{\Delta,l}\,
    \phi^{a,b}_{\Delta+\frac12,l}
    + \gamma^{+2}_{\Delta,l} \, \phi^{a,b}_{\Delta+\frac32,l+1},\\[4mm]
     &2\begin{pmatrix}
    -\sinh\frac{u_1}{2}\ , \ & \sinh\frac{u_2}{2}
    \end{pmatrix} \Psi^{a+\frac14,b-\frac14}_{-,\Delta+1,l} =  \gamma^{-1}_{\Delta,l} \,
    \phi^{a,b}_{\Delta+\frac12,l+1} + \gamma^{-2}_{\Delta,l}\,  \phi^{a,b}_{\Delta+\frac32,l},\\[2mm]
    \label{eq:aux3}
     & 2 \begin{pmatrix}
    \sinh\frac{u_2}{2}\\
    -\sinh\frac{u_1}{2}
    \end{pmatrix} \phi^{a+\frac12,b-\frac12}_{\Delta+1,l} = \tgamma^{1+}_{\Delta,l}\,
    \Psi^{a+\frac14,b-\frac14}_{+,\Delta+\frac12,l}  +  \tgamma^{2+}_{\Delta,l}\,
    \Psi^{a+\frac14,b-\frac14}_{+,\Delta+\frac32,l-1}+ \tgamma^{1-}_{\Delta,l}\,
    {\Psi}^{a+\frac14,b-\frac14}_{-,\Delta+\frac12,l-1}  +  \tgamma^{2-}_{\Delta,l}\,
    {\Psi}^{a+\frac14,b-\frac14}_{-,\Delta+\frac32,l}\ .  
\end{align}
Here
\begin{align}
    & \gamma^{+1}_{\Delta,l} = \frac{\sqrt{2}}{i(-1)^{a+b}}\frac{l+2}{l+1}, \quad
    \gamma^{+2}_{\Delta,l} = \frac{i(-1)^{-a-b}}{2\sqrt{2}} \frac{(4a-2l-2\Delta-1)(2\Delta-3)
    (4b+2l+2\Delta+1)}{(2\Delta-1)(2\Delta+2l+1)(2\Delta+2l+3)},\label{eq:gamma1}\\[2mm]
    & \gamma^{-1}_{\Delta,l} = \frac{2\sqrt{2}}{i(-1)^{a+b}}, \quad \gamma^{-2}_{\Delta,l} =
    \frac{(-1)^{-a-b}}{i\sqrt{2}}\frac{(l+2)(4a+2l-2\Delta+5)(2\Delta-3)(-4b+2l-2\Delta+5)}
    {(l+1)(2l-2\Delta+3)(2l-2\Delta+5)(2\Delta-1)},\label{eq:gamma2}\\[2mm]
    & \tgamma^{1+}_{\Delta,l} = i (-1)^{a+b} 2\sqrt{2}, \quad
    \tgamma^{2+}_{\Delta,l}=\frac{i(-1)^{a+b}}{\sqrt{2}} \frac{ \Delta l (2a+l-\Delta+2)
    (-2b+l-\Delta+2)}{(\Delta-1)(l+1)(l-\Delta+1)(l-\Delta+2)},\label{eq:gamma3}\\[2mm]
    & \tgamma^{1-}_{\Delta,l} = i(-1)^{a+b}\sqrt{2}\frac{l}{l+1},\ \tgamma^{2-}_{\Delta,l}=
    \frac{(-1)^{a+b}}{2\sqrt{2}i}\frac{\Delta(2a-l-\Delta)
    (2b+l+\Delta)}{(\Delta-1)(l+\Delta)(l+\Delta+1)}\ . \label{eq:gamma4}
\end{align}
As far as we are aware, these identities involving scalar and seed conformal blocks are new, although their
form has a clear representation-theoretic origin. Rather than computing the coefficients $\gamma$ in terms of $SO(6)$
Clebsch-Gordan coefficients, they are most easily obtained by picking any point $(u_1,u_2)$ and Taylor-expanding both
sides of the above equations. This produces a system of linear equations and it is clear that going to high enough
order gives sufficiently many equations to fix the eight coefficients $\gamma$. A natural choice for the expansion
point is to take \(u_1\gg u_2\rightarrow\infty \), which corresponds to the OPE limit for the scalar and seed blocks.
\medskip

It is easy to promote the functions $\Psi$ and $\phi$ to eigenfunctions of $H_0$. To this end let us denote the standard
basis of $\mathbb{C}^4$ by $\{e_1,...,e_4\}$. The eigenfunctions of $H_0$ are therefore of the form
\begin{equation}
    \FG_1^0 = \phi^{a,b}_{\Delta+1,l}e_1,\quad \FG_2^0 = \Psi^{a+\frac14,b-\frac14}_{+,\Delta+1,l},\quad
    \FG_3^0 = \Psi^{a+\frac14,b-\frac14}_{-,\Delta+1,l},\quad \FG_4^0 = \phi^{a+\frac12,b-\frac12}_{\Delta+1,l}e_4,
\end{equation}
where it is understood that the two non-zero components of $\FG_2$ and $\FG_3$ are in the space that
is spanned by $e_2$ and $e_3$. We shall work with this notation throughout the remaining part of this
section whenever we write $\Psi_\pm$. The solution of the eigenvalue problem for $H$, which can be obtained
by a finite perturbation of $\FG^0_i$, will be denoted by $\FG_i$, as before. With the help of the explicit
expressions for $\FG^0_i$, the formula \eqref{nilpotent} for the nilpotent potential and our auxiliary formulas
(\ref{eq:aux1})-(\ref{eq:aux3}) one finds\footnote{The tensor product of any given finite dimensional \(SO(d,2)\) representation \(T_\nu\) labelled by a Young tableau \(\nu\) and the induced representation \(\pi_{\Delta,\mu}\) 
can be decomposed into a finite sum of induced representations (e.g. see \cite{Karateev:2017jgd}) as
\begin{gather}
T_\nu \otimes \pi_{\Delta,\mu}=\bigoplus\limits_{i=-j}^j \bigoplus_{\lambda\in \nu^i\otimes\mu}\pi_{\Delta+i,\lambda} \label{FinDimTimesInduced}
\end{gather}
where indices \((i,\nu^i)\) are defined through the decomposition of the \(SO(d+2)\) representation \(\nu\) with respect to its \(SO(2)\times SO(d)\) subgroup and enumerate a (semi) integer \(SO(2)\) conformal weight \(i\)  
along with an \(SO(d)\) Young tableau \(\nu^i\).

Bosonic conformal blocks are particular matrix elements of some representation $\pi_{\Delta,\mu}$ of $G_{(0)}$, \cite{Schomerus:2016epl}. In the course of perturbation theory they are multiplied by matrix elements of the fundamental representation $\pi=\pi_f$ of $G_{(0)}$. Therefore, the bosonic blocks that appear in the $n$-th order of the perturbation theory are matrix elements of $\pi_{\Delta,\mu}\otimes \pi_f^{\otimes n}$. This fixes the functional form of our solutions.}
\begin{align}
    & \FG_1 = \FG^0_1 = \phi^{a,b}_{\Delta+1,l}e_1, \label{eq:F1} \\[2mm]
    & \FG_2 = \Psi^{a+\frac14,b-\frac14}_{+,\Delta+1,l}+\left(\c^{+1}_{\Delta,l}\, \phi^{a,b}_{\Delta+\frac{1}{2},l}
    +\c^{+2}_{\Delta,l}\, \phi^{a,b}_{\Delta+\frac{3}{2},l+1}\right)e_1, \label{eq:F2} \\[2mm]
    & \FG_3 = \Psi^{a+\frac14,b-\frac14}_{-,\Delta+1,l}+ \left(\c^{-1}_{\Delta,l}\,
    \phi^{a,b}_{\Delta+\frac12,l+1}+ \c^{-2}_{\Delta,l}\, \phi^{a,b}_{\Delta+\frac{3}{2},l}\right)e_1, \label{F3ansatz}\\[2mm]
    & \FG_4  = \phi^{a+\frac12,b-\frac12}_{\Delta+1,l}e_4 + \tc^{1+}_{\Delta,l}\, \Psi^{a+\frac14,b-\frac14}_{+,\Delta+\frac12,l}
    + \tc^{2+}_{\Delta,l}\, \Psi^{a+\frac14,b-\frac14}_{+,\Delta+\frac32,l-1}  + \tc^{1-}_{\Delta,l}\,
    \Psi^{a+\frac14,b-\frac14}_{-,\Delta+\frac12,l-1}  + \tc^{2-}_{\Delta,l}\,
    {\Psi}^{a+\frac14,b-\frac14}_{-,\Delta+\frac32,l} + \nonumber \\[2mm]
    & \hspace*{2cm} +
    \left(k^{00}_{\Delta,l}\, \phi^{a,b}_{\Delta,l} +k^{01}_{\Delta,l}\, \phi^{a,b}_{\Delta+1,l+1} +
    k^{10}_{\Delta,l}\, \phi^{a,b}_{\Delta+1,l-1}+k^{11}_{\Delta,l}\, \phi^{a,b}_{\Delta+2,l}  \right)e_1\ . \label{eq:F4}
\end{align}
The construction of these four solutions requires increasing orders of perturbation theory. The solution $\FG_1$
is obviously obtained in zeroth order and equal to the scalar bosonic block $\FG_1^0$. The second and the third
solution $\FG_2$ and $\FG_3$ are obtained at first order while the last solution $\FG_4$ required to go to second
order. In order to complete the solution, we just need to spell out the various coefficients. Those that arose from
first order perturbation theory are directly related to the coefficients $\gamma$ we introduced in eqs.\
(\ref{eq:gamma1})-(\ref{eq:gamma4}) as,
\begin{equation}\label{eq:c}
    c^{\pm i} = \frac{\gamma^{\pm i}}{\frac{3}{16}(2r+3) + C^{\sc}_{\pm i} - C^{\seed}_{\Delta+1,l}}\ \quad , \quad \quad
    c^{i \pm} = \frac{\gamma^{i \pm}}{\frac{3}{16}(2r+1)+C^{\seed}_{i\pm }-C^{\sc}_{\Delta+1,l}}
\end{equation}
where
\begin{equation}
\begin{pmatrix} C^\s_{+1} & C^\s_{+2} \\ C^\s_{-1} & C^\s_{-2} \end{pmatrix} =
\begin{pmatrix} C^\s_{\Delta + \frac12,l} & C^\s_{\Delta+ \frac32,l+1} \\
C^\s_{\Delta+ \frac12, l+1} & C^\s_{\Delta+\frac32, l} \end{pmatrix}\ ,
\quad
\begin{pmatrix} C^\s_{1+} & C^\s_{1-} \\ C^\s_{2+} & C^\s_{2-} \end{pmatrix} =
\begin{pmatrix} C^\s_{\Delta + \frac12,l} & C^\s_{\Delta+ \frac12, l-1} \\
C^\s_{\Delta + \frac32, l-1} & C^\s_{\Delta+\frac32, l} \end{pmatrix}\ ,
\end{equation}
for $\s = \sc,\seed$. The second order coefficients in the second line of $\FG_4$ involve products of the coefficients $\gamma$
and they read as follows,
\begin{align} \label{eq:k1}
    & k^{00}_{\Delta,l} = \frac{\gamma^{1+}_{\Delta,l}\gamma^{+1}_{\Delta-\frac12,l}+\gamma^{1-}_{\Delta,l}
    \gamma^{-1}_{\Delta-\frac12,l-1}}{\frac34(r+1)-C^{\sc}_{\Delta+1,l}+C^{\sc}_{\Delta,l}}\ ,\quad \quad \quad
    k^{01}_{\Delta,l} = \frac{\gamma^{1+}_{\Delta,l}\gamma^{+2}_{\Delta-\frac12,l} + \gamma^{2-}_{\Delta,l}
    \gamma^{-1}_{\Delta+\frac12,l}} {\frac34(r+1)-C^{\sc}_{\Delta+1,l}+C^{\sc}_{\Delta+1,l+1}},\\[2mm]
    & k^{10}_{\Delta,l} = \frac{\gamma^{2+}_{\Delta,l}\gamma^{+1}_{\Delta+\frac12,l-1} + \gamma^{1-}_{\Delta,l}
    \gamma^{-2}_{\Delta-\frac12,l-1}}{\frac34(r+1)-C^{\sc}_{\Delta+1,l}+C^{\sc}_{\Delta+1,l-1}}\ ,\quad \quad
    k^{11}_{\Delta,l} = \frac{\gamma^{2+}_{\Delta,l}\gamma^{+2}_{\Delta+\frac12,l-1} + \gamma^{2-}_{\Delta,l}
    \gamma^{-2}_{\Delta+\frac12,l}}{\frac34(r+1)-C^{\sc}_{\Delta+1,l}+C^{\sc}_{\Delta+2,l}}\ . \label{eq:k2}
\end{align}
Clearly, Hamiltonians $H_0$ and $H$ have the same spectrum - the eigenvalue of $\FG_i$ equals that of $\FG^0_i$.
The dependence of these eigenvalues on the weight $\Delta$ and spin $l$ of the intermediate field can be read off
from (\ref{eigenvalues-1})-(\ref{eigenvalues-2}),
\begin{align}\label{eigenvalues-3}
    & C_2 = C_3 = \frac14 \Delta(\Delta-2) + \frac14 l(l+3) - \frac{3}{16}(r+1)^2 +\frac38,\\[2mm]
    & C_1 = \frac14 \Delta(\Delta-2) + \frac14 l(l+2) -\frac{3}{16}(r+2)^2,\ C_4 =
    \frac14 \Delta(\Delta-2) + \frac14 l(l+2) -\frac{3}{16}r^2 \ .\label{eigenvalues-4}
\end{align}
This completes our discussion of the solution to the eigenvalue problem of $H$ and hence our construction
of the superconformal blocks for the four-point functions under consideration. Let us stress again that all
these results are completely explicit, with explicitly known coefficients $c,k$ in eqs.\ \eqref{eq:c},
\eqref{eq:k1}, \eqref{eq:k2} and explicitly known functions $\phi$ and $\Psi_\pm$, see the appendix E.

\subsection{Assembling the pieces: The crossing equations}

In this last subsection, we will derive the crossing symmetry equations for the four-point function
\eqref{4-point-function} with identical long operators $\mathcal{O}_2 = \mathcal{O}_4 \equiv \mathcal{R}$
of vanishing $R$-charge. In order to achieve this, we first need to associate eigenfunctions of $H$ found
in the previous subsection with superconformal blocks representing the propagation of superconformal
representations in the operator products $\bar\varphi\times R$ and $\varphi\times R$.

There are four possible kinds of representations appearing in these operator products. Their quantum
numbers are, \cite{Li:2017ddj}
\begin{equation}
    \rho_1 = (j,j)^\Delta_{-r-2}, \quad \rho_2 = (j,j+1/2)^\Delta_{-r-1}, \quad
    \rho_3 = (j+1/2,j)^\Delta_{-r-1}, \quad \rho_4 = (j,j)^\Delta_{-r}\ .
\end{equation}
We will write $l = 2j$. One observes that the eigenvalues $C_i$ written in (\ref{eigenvalues-3})-(\ref{eigenvalues-4})
coincide with the values of the quadratic Casimir in representations $\rho_i$ (see the appendix D). Therefore,  we can
identify the propagation of the operator labelled by $\rho_i$ with the superconformal block $\FG_i$ from above.

There still remains the question of normalisation of these blocks. To settle it, let us focus on the $e_1$-component
of the four solutions. From the explicit formulas of the previous section, one can directly verify that
\begin{align}
    & \frac{c^{+2}_{\Delta,l}}{c^{+1}_{\Delta,l}} = -\frac{\hat c_2}{\hat c_1},
    \quad \frac{c^{-2}_{\Delta,l}}{c^{-1}_{\Delta,l}} = -\frac{\check c_2}{\check c_1},
    \quad \frac{k^{01}_{\Delta,l}}{k^{00}_{\Delta,l}} = -\bar c_1,
    \quad \frac{k^{10}_{\Delta,l}}{k^{00}_{\Delta,l}} = -\bar c_2,
    \quad \frac{k^{11}_{\Delta,l}}{k^{00}_{\Delta,l}} = \bar c_1 \bar c_2\ .
\end{align}
Here, the coefficients $\hat c_i, \check c_i, \bar c_i$ are taken from \cite{Li:2017ddj},
with the substitution $l\xrightarrow{}l+1$ in coefficients $\check c_i$ in order to
synchronise conventions. We will only need explicit formulas for two of these coefficients
below,
\begin{equation} \label{eq:Lic}
\hat c_1 = \frac{l+2}{(l+1)(2(\Delta - l - \Delta_\varphi) -3)} \ , \quad
\check c_1 = \frac{1}{2(\Delta + l -\Delta_\varphi) + 3} \ .
\end{equation}
After stripping of a conventional prefactor, the four-point function in \cite{Li:2017ddj}
decomposes over superconformal blocks as
\begin{equation}\label{eq:Lig}
    g(z_i) = \sum |c_{\bar\varphi R(\bar Q^2\mathcal{O}_l})|^2 \mathcal{G}(z_i) +
    \sum |c_{\bar\varphi R(\bar Q\mathcal{O})_l}|^2 \mathcal{\hat G}(z_i) +
    \sum |c_{\bar\varphi R(\bar Q\mathcal{O})_l}|^2 \mathcal{\check G}(z_i) +
    \sum |\bar c_{\bar\varphi R\mathcal{O}_l}|^2 \mathcal{\bar G}(z_i)\ .
\end{equation}
Here $\mathcal{G},\mathcal{\hat G}, ...$ are sums of Dolan-Osborn scalar conformal blocks
with appropriate coefficients, see eqs. \textit{(2.32),(2.25),(2.29),(2.19)} of
\cite{Li:2017ddj} for details. The Calogero-Sutherland scalar blocks are related to
Dolan-Osborn ones by $\phi^{a,b}_{\Delta,l} = \Lambda^{a,b}(z_i)g^{a,b}_{\Delta,l}$,
where $\Lambda^{a,b}$ is an explicit function of cross ratios whose precise expression
we will not need, see  \cite{Isachenkov:2016gim}. The relation of the zero component
of our function $\FG$ to the function $g$ defined in eq.\ \eqref{eq:Lig} mimics that
between the blocks, $\FG^0 =\Lambda^{a,b}(z_i) g$. Therefore, the function $\FG$ on
the abelian group $A$ that encodes the correlation function \eqref{4-point-function}
in our approach decomposes over the superconformal blocks $\FG_j$, $j=1,...,4$, we
determined on the previous subsection with coefficients
\begin{align}\label{eq:block}
    \FG(\alpha_i) &= \sum |c_{\bar\varphi R(\bar Q^2\mathcal{O})_l}|^2 \FG_1(\alpha_i)
    + \sum \frac{\hat c_1}{c^{+1}_{\Delta,l}}
    |c_{\bar\varphi R(\bar Q\mathcal{O})_l}|^2 \FG_2(\alpha_i)\\[2mm]
    & + \sum\frac{\check c_1}{c^{-1}_{\Delta,l}} |c_{\bar\varphi R(\bar Q\mathcal{O})_l}|^2 \FG_3(\alpha_i) +
    \sum\frac{1}{k^{00}_{\Delta,l}} |\bar c_{\bar\varphi R\mathcal{O}_l}|^2 \FG_4(\alpha_i)\ .
\end{align}
We write interchangeably $\alpha=\alpha_1$ and $\alpha^\ast=\alpha_2$. The relation of
$\alpha_i$ with the cross ratios was stated in eq.\ \eqref{eq:alpha}. Combining the block
decomposition \eqref{eq:block} of the correlation function $\FG$ with the crossing symmetry
relation between $s-$ and $t-$channel,
\begin{equation}
    \FG(1/\alpha,1/\alpha^\ast) = M_{st}(\alpha,\alpha^\ast) \FG(\alpha,\alpha^\ast)\
\end{equation}
with $M_{st}$ as defined in eq.\ \eqref{eq:Mst} we have thereby derived the crossing relations
we anticipated in eq.\ \eqref{eq:crossing} of the introduction with coefficients given by
\begin{equation} \label{eq:gamma}
\hat{\gamma} = \frac{\hat{c}_1}{c^{+1}_{\Delta,l}}\ , \quad \check{\gamma} =
\frac{\check{c}_1}{c^{-1}_{\Delta,l}} \, \quad \bar{\gamma} = \frac{1}{k^{00}_{\Delta,l}} \ .
\end{equation}
with the coefficients $\hat{c}_1$ and $\check{c}_1$ defined in eq.\ \eqref{eq:Lic}, as
well as the coefficients $c^{\pm 1}_{\Delta,l}$ and $k^{00}_{\Delta,l}$ we introduced in
eqs.\ \eqref{eq:c} and \eqref{eq:k1}, respectively. All these expressions are rational
functions of the conformal weight $\Delta$ and the spin $l$. Thereby we have completed
our derivation of the crossing equation \eqref{eq:crossing}.

\section{Conclusions}

In this work we constructed superconformal blocks and crossing symmetry equations for the four-point function of two long, one chiral and one anti-chiral scalar operator in $4$-dimensional $\mathcal{N}=1$ SCFTs. The two ingredients that went into the derivation, namely the construction of superblocks $\FG_j$ and of the crossing factor $M_{st}$ applied general techniques developed in \cite{Buric:2019rms} and \cite{Buric:2020buk}. This is indeed the first time a full set of long multiplet bootstrap equations has been written down in dimension higher than two. In the course of solving the Casimir differential equations for the relevant superconformal blocks, we also derived 
certain relations between scalar and seed bosonic conformal blocks which seem to be new.

As should be clear from the previous sections, our methods are by no means restricted to the above class of correlators. They are completely algorithmic and can be applied to a wide class of correlation functions involving long multiplets in SCFTs with superconformal symmetry of type I. Let us mention some directions that can be pursued in the future.

The approach to superconformal partial waves we developed in \cite{Buric:2019rms} works particular well for long multiplets. While some shortening conditions may be implemented before even writing down the Casimir differential equations, others can only be implemented once the solutions to these equations have been constructed. Which case occurs depends on how compatible the specific set of shortening conditions is with our choice or Cartan coordinates, see \cite{Buric:2019rms} for details. The four-point functions \eqref{4-point-function} we addressed here involve two BPS operators. Their shortening conditions belong to the first class that are easy to implement before writing down the equations. While the Casimir equations for the correlation function of four long multiplets $\mathcal{R}$ turn out to possess $36$ components, the shortening conditions of the chiral and anti-chiral field reduces the operators to a $4 \times 4$ matrix system. These statements are particular to $\mathfrak{sl}(4|1)$, but the same simplifications occur for any superconformal algebra of type I. Also, the simplifications in the crossing factor that we saw in section 2.4 have certain universality and appear for other algebras of the infinite family $\mathfrak{sl}(m|\mathcal{N})$. Therefore, correlators of the type $(\ref{4-point-function})$ in a large class of SCFTs are very well-suited to the Calogero-Sutherland gauge.

In order to perform a numerical long multiplet bootstrap analysis for $\mathcal{N}=1$ SCFTs in $d=4$, one would need to include the four-point function of the superfield $\mathcal{R}$ in order to exploit positivity. This four-point function can be addressed with the same techniques as used in this work. Writing down the Casimir equations is actually not that difficult. Finding solutions meets two technical challenges. On the one hand, the unperturbed Schr\"odinger problem for $H_0$ now involves a larger set of 4-dimensional spinning blocks, including both higher seed and non-seed blocks. Of course, all of these are known in the literature and can be mapped to the Calogero-Sutherland gauge using results of \cite{Buric:2019dfk}. On the other hand, the nilpotent perturbation theory has to be taken to a higher order before it truncates. This requires a vast extension of the relations (\ref{eq:aux1})-(\ref{eq:aux3}). While all this is in principle straightforward it eventually results in relatively large sums of bosonic blocks. The superprimary component of the associated crossing symmetry equation has been worked out in \cite{Li:2017ddj}. So, the main task that remains is to construct the other $35$ components of this system of equations. Given that one would expect significantly stronger constraints on the dynamics of $\mathcal{N}=1$ SCFTs it may be well worth the effort to carry this out explicitly.   
 
Similar comments apply to the important question of bootstrapping the stress tensor multiplet in four-dimensional $\mathcal{N}=2$ SCFTs. It would require significant technical effort, but our method does apply to this case as well. Let us note that, while the method follows a clear algorithm, not all of its steps have been automatised by computer programs yet. Even without this, we were able to treat the highly non-trivial correlator $(\ref{4-point-function})$ because symmetries are exploited to such an extent that computations in the end become quite simple. Therefore, the idea that the fully automatised program will turn the difficult problems mentioned above into tractable or even routine ones holds some promise.

In another direction, it would be interesting to study from the Calogero-Sutherland side the observed relations between three type I superconformal algebras in dimensions two, three and four, namely $\mathfrak{sl}(2|1)\oplus\mathfrak{sl}(2|1)$, $\mathfrak{osp}(2|4)$ and $\mathfrak{sl}(4|1)$. These Lie superalgebras have been observed to possess similar properties which allow for them to be studied in a somewhat uniform manner, \cite{Bobev:2015jxa}. We have already analysed $\mathfrak{sl}(2|1)$ in \cite{Buric:2019rms} and $\mathfrak{sl}(4|1)$ here. The remaining algebra, $\mathfrak{osp}(2|4)$, which is the only type I superconformal algebra in three dimensions, is in many respect similar to $\mathfrak{sl}(4|1)$ - it has the same number of odd generators and admits the same kind of short representations. Its analysis would thus be analogous to the one presented here, with one difference: bosonic blocks in three dimensions are more complicated than in four and cannot be written as finite sums of Gauss' hypergeometric functions. It would be interesting to compute the Calogero-Sutherland model associated to $\mathfrak{osp}(2|4)$ and compare it to the ones found here and in \cite{Buric:2019rms}. 

As a final comment, the harmonic analysis approach can be extended to defect SCFTs. The corresponding bosonic theory of defect conformal blocks and tensor structures has been constructed in \cite{defects}. Similarly to the non-defect case, the theory admits a natural generalisation to include supersymmetry. Since both the theory of \cite{Buric:2019rms,Buric:2020buk} and that of \cite{defects} are highly structured and have various advantages over the conventional approaches, there is reason to believe that the same will be true for the combination of the two. Defects in superconformal theories have a rich structure and diverse applications, so exploring this direction further certainly seems worthwhile.
\bigskip 
\bigskip 

\noindent
{\bf Acknowledgements:} We thank Aleix Gimenez-Grau, Misha Isachenkov, Denis Karateev, Madalena Lemos, Pedro Liendo, 
Junchen Rong, Andreas Stergiou and Philine van Vliet for comments and fruitful discussion. The work of ES is supported by Ministry of Science and Higher Education of the Russian Federation, agreement \textnumero \ 075-15-2019-1620 and by ERC grant 648630 IQFT.  VS and IB acknowledge support by the Deutsche Forschungsgemeinschaft (DFG, German Research 
Foundation) under Germany’s Excellence Strategy – EXC 2121 ,,Quantum Universe'' – 390833306.

\appendix

\section{Superconformal Algebras of Type I}

In this appendix we define what is meant by a superconformal algebra and introduce types I and II. While some of the discussion of the main text applies equally well to both types, the construction of Casimir equations relies on the algebra being of type I.

Let $\mathfrak{g} = \mathfrak{g}_{(0)} \oplus \mathfrak{g}_{(1)}$ be a finite-dimensional Lie superalgebra. We say that $\mathfrak{g}$ is a superconformal algebra if its even part $\mathfrak{g}_{(0)}$ contains the conformal Lie algebra $\mathfrak{so}(d+1,1)$ as a direct summand and the odd part $\mathfrak{g}_{(1)}$ decomposes as a direct sum of spinor representations of $\mathfrak{so}(d)\subset\mathfrak{so}(d+1,1)$ under the adjoint action.

If this is the case, we denote the dilation generator of the bosonic conformal Lie algebra by $D$. Eigenvalues with respect to $\text{ad}_D$ give a decomposition of $\mathfrak{g}$ into the sum of eigenspaces
\begin{equation}
\mathfrak{g} = \mathfrak{g}_{-1}\oplus\mathfrak{g}_{-1/2} \oplus \mathfrak{g}_0\oplus\mathfrak{g}_{1/2}\oplus\mathfrak{g}_{1} = \mathfrak{g}_{-1} \oplus \mathfrak{s} \oplus \mathfrak{k} \oplus \mathfrak{q}\oplus \mathfrak{g}_{1}\ .
\end{equation}
The even part of $\mathfrak{g}$ is composed of $\mathfrak{g}_{\pm 1}$ and $\mathfrak{k}$ where $\mathfrak{g}_{-1}={\mathfrak{n}}$ contains the generators $K_\mu$ of special conformal
transformations while $\mathfrak{g}_{1} = \mathfrak{n}$ is spanned by translations $P_\mu$. Dilations, rotations and internal symmetries make up $$ \mathfrak{k} = \mathfrak{so}(1,1) \oplus \mathfrak{so}(d) \oplus \mathfrak{u} \ . $$ Generators of $\mathfrak{g}_{\pm1/2}$, are supertranslations $Q_\alpha$ and super special conformal transformations $S_\alpha$. We shall also denote these summands as $\mathfrak{s}
= \mathfrak{g}_{-1/2}$ and $\mathfrak{q} = \mathfrak{g}_{1/2}$. All elements of non-positive degree make up a subalgebra $\mathfrak{p}$ of $\mathfrak{g}$ that will be referred to as the  parabolic subalgebra
\begin{equation}
     \mathfrak{p} = \mathfrak{g}_{-1} \oplus \mathfrak{g}_{-1/2} \oplus \mathfrak{g}_0\ .
\end{equation}
There is a unique (connected) corresponding subgroup $P\subset G$ such that $\mathfrak{p} = \Lie(P)$. The superspace can be identified with the supergroup of translations and supertranslations. It is defined as the homogeneous space $M = G/P$.

The above structure is present in any superconformal algebra. In this work, we shall mainly consider those $\mathfrak{g}$ which satisfy an additional condition of being of type I. This means that the
odd subspace decomposes as a direct sum of two irreducible representations of $\mathfrak{g}_{(0)}$ under the adjoint action
\begin{equation}
\mathfrak{g}_{(1)} = \mathfrak{g}_+ \oplus \mathfrak{g}_-\ .
\end{equation}
The two modules $\mathfrak{g}_\pm$ are then necessarily dual to each other and further satisfy
\begin{equation}
    \{\mathfrak{g}_\pm , \mathfrak{g}_\pm\} = 0 \ .
\end{equation}
In addition, the bosonic algebra assumes the form
\begin{equation}
\mathfrak{g}_{(0)} = [\mathfrak{g}_{(0)},\mathfrak{g}_{(0)}] \oplus \mathfrak{u}(1)\ .
\end{equation}
The $\mathfrak{u}(1)$ summand is a part of the internal symmetry algebra. Its generators will be denoted by $R$. All elements in $\mathfrak{g}_+$ possess the same $R$-charge. The same is true for the
elements of $\mathfrak{g}_-$, but the $R$-charge of these elements has the opposite value. Elements in the even subalgebra $\mathfrak{g}_{(0)}$, on the other hand, commute with $R$.

Let us denote the intersections of the subspaces $\mathfrak{q}$ and $\mathfrak{s}$ with $\mathfrak{g}_\pm$ by
\begin{equation} \label{eq:split}
\mathfrak{q}_\pm = \mathfrak{q}\cap\mathfrak{g}_{\pm} \quad,\ \mathfrak{s}_\pm =\mathfrak{s}\cap\mathfrak{g}_\pm\ .
\end{equation}
The subspaces $\mathfrak{q}_\pm$ and $\mathfrak{s}_\pm$ do not carry a representation of $\mathfrak{g}_{(0)}$, but they do carry a representation of $\mathfrak{k}$. This also
means that in type I superconformal algebras, the action of $\mathfrak{k}$ on super-translations decomposes into two or more irreducible representations. It turns out that
\begin{equation}
    \text{dim}(\mathfrak{q}_\pm) = \text{dim}(\mathfrak{s}_\pm)  = \text{dim}(\mathfrak{g}_{(1)})/4\ .
\end{equation}
The full list of type I superconformal algebras, which follows directly from Kac's classification \cite{Kac:1977em}, is
\begin{equation}
    \mathfrak{sl}(2|\mathcal{N}),\ \mathfrak{sl}(2|\mathcal{N}_1)\oplus\mathfrak{sl}(2|\mathcal{N}_2)\ \mathfrak{psl}(2|2),\ \mathfrak{osp}(2|4),\ \mathfrak{sl}(4|\mathcal{N}),\ \mathfrak{psl}(4|4)\ .
\end{equation}
The presented list is that of complexified Lie superalgebras - for different spacetime signatures one considers their various real forms.

\section{Conventions for $\mathfrak{sl}(2m|\mathcal{N})$}

In this appendix we collect our conventions for the class of Lie superalgebras $\mathfrak{g} = \mathfrak{sl}(2m|\mathcal{N})$. Before doing that, let us state conventions regarding spinors that are used throughout the main text. Greek indices from the middle of the alphabet $\mu, \nu...=1,...,4$ are Euclidean spacetime indices and are raised and lowered using the flat Euclidean metric $g_{\mu \nu} = \delta_{\mu\nu}$.

The undotted and dotted Greek indices from the beginning of the alphabet $\alpha,\dot \alpha... = 1,2$ are labelling vectors in the $(1/2,0)$ and $(0,1/2)$ representations of $\mathfrak{so}(4)$, respectively. They are raised and lowered using the Levi-Civita symbol
\begin{equation}
   \psi_\alpha = \varepsilon_{\alpha\beta}\psi^\beta,\quad  \varepsilon_{\alpha\beta} = \begin{pmatrix}
     0 & -1\\
     1 & 0
    \end{pmatrix}\ .
\end{equation}
The same rule holds for dotted indices. To convert a vector index into a pair of a fundamental and and an anti-fundamental index, we make use of the matrices
\begin{equation}
    (\gamma_\mu)^{\dot\alpha}_{\ \alpha} = (-\sigma_3,-i I_2,\sigma_1,-\sigma_2) \quad \text{i.e.} \quad x^{\dot\alpha}_{\ \alpha} = \begin{pmatrix}
     -x_1 - i x_2 & x_3+i x_4\\
     x_3 - i x_4 & x_1-i x_2
    \end{pmatrix}\ .
\end{equation}
We now give the bracket relations in $\mathfrak{g}$. The even subalgebra $\mathfrak{g}_{(0)}$ has brackets
\begin{align}
	&[D, P_{\dot\alpha}^{\ \beta}] = P_{\dot\alpha}^{\ \beta}\ , \ [D, K_{\alpha}^{\ \dot\beta} ]=- K_{\alpha}^{\ \dot\beta}, \\[2mm]
	&[M_{\alpha}^{\ \beta},P_{\dot\gamma}^{\ \delta}] = \frac12 \delta_{\alpha}^{\ \beta} P_{\dot\gamma}^{\ \delta} - \delta_{\alpha}^{\ \delta}P_{\dot\gamma}^{\ \beta},\ [M_{\dot\alpha}^{\ \dot\beta},P_{\dot\gamma}^{\ \delta}] = -\frac12 \delta_{\dot\alpha}^{\ \dot\beta} P_{\dot\gamma}^{\ \delta} + \delta_{\dot\gamma}^{\ \dot\beta}P_{\dot\alpha}^{\ \delta},  \\[2mm]
	&[M_{\alpha}^{\ \beta},K_{\gamma}^{\ \dot\delta}] = -\frac12 \delta_{\alpha}^{\ \beta} K_{\gamma}^{\ \dot\delta} + \delta_{\gamma}^{\ \beta}K_{\alpha}^{\ \dot\delta},\ [M_{\dot\alpha}^{\ \dot\beta},K_{\gamma}^{\ \dot\delta}] = \frac12 \delta_{\dot\alpha}^{\ \dot\beta} K_{\gamma}^{\ \dot\delta} - \delta_{\dot\alpha}^{\ \dot\delta}K_{\gamma}^{\ \dot\beta},  \\[2mm]
    & [M_{\dot\alpha}^{\ \dot\beta}, M_{\dot\gamma}^{\ \dot\delta}] = \delta_{\dot\gamma}^{\ \dot\beta} M_{\dot\alpha}^{\ \dot\delta} - \delta_{\dot\alpha}^{\ \dot\delta} M_{\dot\gamma}^{\ \dot\beta},\ [M_{\alpha}^{\ \beta}, M_{\gamma}^{\ \delta}] = \delta_{\gamma}^{\ \beta} M_{\alpha}^{\ \delta} - \delta_{\alpha}^{\ \delta} M_{\gamma}^{\ \beta}, \\[2mm]
	& [K_{\alpha}^{\ \dot\beta},P_{\dot\gamma}^{\ \delta}] =  \delta_{\dot\gamma}^{\ \dot\beta}M_{\alpha}^{\ \delta} - \delta_{\alpha}^{\ \delta}M_{\dot\gamma}^{\ \dot\beta} - 2 \delta_{\dot\gamma}^{\ \dot\beta}\delta_{\alpha}^{\ \delta} D\ .
\end{align}
Next, the brackets between even and odd generators read
\begin{align}
    & [R,Q_{\dot\alpha}^{\ J}] = Q_{\dot\alpha}^{\ J},\ [R,Q_{I}^{\ \beta}] = -Q_{I}^{\ \beta},\ [R,S_{\alpha}^{\ J}] = S_{\alpha}^{\ J},\ [R,S_{I}^{\ \dot\beta}] = -S_{I}^{\ \dot\beta}, \\[2mm]
	& [D,Q_{\dot\alpha}^{\ J}] = \frac12 Q_{\dot\alpha}^{\ J},\ [D,Q_{I}^{\ \beta}] = \frac12 Q_{I}^{\ \beta},\ [D,S_{\alpha}^{\ J}] = -\frac12 S_{\alpha}^{\ J},\ [D,S_{I}^{\ \dot\beta}] = -\frac12 S_{I}^{\ \dot\beta},\\[2mm]
	& [M_{\alpha}^{\ \beta},Q_{K}^{\ \delta}] = \frac12 \delta_{\alpha}^{\ \beta} Q_{K}^{\ \delta} - \delta_{\alpha}^{\ \delta}Q_{K}^{\ \beta},\ [M_{\dot\alpha}^{\ \dot\beta},Q_{\dot\gamma}^{\ L}] = -\frac12 \delta_{\dot\alpha}^{\ \dot\beta} Q_{\dot\gamma}^{\ L} + \delta_{\dot\gamma}^{\ \dot\beta}Q_{\dot\alpha}^{\ L},  \\[2mm]
	& [M_{\alpha}^{\ \beta},S_{\gamma}^{\ L}] = -\frac12 \delta_{\alpha}^{\ \beta} S_{\gamma}^{\ L} + \delta_{\gamma}^{\ \beta}S_{\alpha}^{\ L},\ [M_{\dot\alpha}^{\ \dot\beta},S_{K}^{\ \dot\delta}] = \frac12 \delta_{\dot\alpha}^{\ \dot\beta} S_{K}^{\ \dot\delta} - \delta_{\dot\alpha}^{\ \dot\delta}S_{K}^{\ \dot\beta},  \\[2mm]
	& [P_{\dot\alpha}^{\ \beta},S_{\gamma}^{\ L}] = \delta_{\gamma}^{\ \beta}Q_{\dot\alpha}^{\ L},\ [P_{\dot\alpha}^{\ \beta},S_{K}^{\ \dot\alpha}] = -\delta_{\dot\alpha}^{\ \dot\delta}Q_{K}^{\ \beta},\ [K_{\alpha}^{\ \dot\beta},Q_{\dot\gamma}^{\ L}] = \delta_{\dot\gamma}^{\ \dot\beta}S_{\alpha}^{\ L},\ [K_{\alpha}^{\ \dot\beta},Q_{K}^{\ \delta}] = -\delta_{\alpha}^{\ \delta}S_{K}^{\ \dot\beta}\ .
\end{align}
Finally, we give the brackets between odd generators
\begin{align}
	& \{ Q_{\dot\alpha}^{\ J},Q_{I}^{\ \beta}\} = \delta_{I}^{\ J} P_{\dot\alpha}^{\ \beta},\ \{ S_{\alpha}^{\ J},S_{I}^{\ \dot\beta}\} = \delta_{I}^{\ J} K_{\alpha}^{\ \dot\beta}, \\[2mm]
	& \{ Q_{\dot\alpha}^{\ J}, S_{I}^{\ \dot\beta}\} = \delta_{I}^{\ J}M_{\dot\alpha}^{\ \dot\beta} + \delta_{\dot\alpha}^{\ \dot\beta} R_{I}^{\ J} + \delta_{I}^{\ J}\delta_{\dot\alpha}^{\ \dot\beta}(aD+bR),\\[2mm]
	& \{ Q_{I}^{\ \beta}, S_{\alpha}^{\ J}\} = \delta_{I}^{\ J}M_{\alpha}^{\ \beta} + \delta_{\alpha}^{\ \beta} R_{I}^{\ J} + \delta_{I}^{\ J}\delta_{\alpha}^{\ \beta}(cD+dR)\ .
\end{align}
Throughout the text, we use the fundamental, $2m+\mathcal{N}$-dimensional representation of $\mathfrak{g}$. In this representation, the generators are given by
\begin{equation}
  D =\frac12\text{diag}(I_m,-I_m,0),\quad R = \frac{1}{\mathcal{N}-2m} \text{diag}(\mathcal{N}I_m,\mathcal{N}I_m,2m I_{\mathcal{N}})\ .
\end{equation}
\begin{align}
    & M_{\dot1}^{\ \dot2} = E_{\dot 1}^{\ \dot2},\quad M_{\dot2}^{\ \dot1} = E_{\dot 2}^{\ \dot1},\quad  M_{\dot1}^{\ \dot1} = -  M_{\dot2}^{\ \dot2} = \frac12 (E_{\dot 1}^{\ \dot1} - E_{\dot 2}^{\ \dot2}),\\
    & M_{1}^{\ 2} = E_{1}^{\ 2},\quad M_{2}^{\ 1} = E_{2}^{\ 1},\quad  M_{1}^{\ 1} = - M_{2}^{\ 2} = \frac12 (E_{1}^{\ 1} - E_{2}^{\ 2})\ .
\end{align}
\begin{equation}
    P_{\dot\alpha}^{\ \beta} = E_{\dot\alpha}^{\ \beta},\quad Q_{\dot\alpha}^{\ J} = E_{\dot\alpha}^{\ J},\quad Q_I^{\ \beta} = E_I^{\ \beta},\quad K_{\alpha}^{\ \dot\beta} = E_{\alpha}^{\ \dot\beta},\quad S_{\alpha}^{\ J} = E_{\alpha}^{\ J},\quad S_{I}^{\ \dot\beta} = E_I^{\ \dot\beta}\ .
\end{equation}
\begin{equation}
    R_I^{\ J} = E_I^{\ J}, \quad R_I^{\ I} = \frac12 (E_I^{\ I} - E_{I+1}^{\ I+1})\ .
\end{equation}
The standard differential operators representing the action of translation and supertranslation generators on the superspace are
\begin{align}
    p_{\dot\alpha}^{\ \alpha} = \partial_{\dot\alpha}^{\ \alpha},\quad q_{\dot\alpha}^{\ I} = \partial_{\dot\alpha}^{\ I} - \frac12\bar\theta^I_{\ \beta}\partial_{\dot\alpha}^{\ \beta},\quad q_I^{\ \alpha} = \partial_{I}^{\ \alpha}-\frac12\theta^{\dot\beta}_{\ I}\partial_{\dot\beta}^{\ \alpha}\ .
\end{align}

\section{Conventions for $\mathfrak{sl}(4|1)$}

\subsection{Killing form}

As the standard basis of $\mathfrak{g}_{(0)}$ we take the generators in the order
\begin{equation}
    \{X^a\} = \{ R,D,M_{\dot1}^{\ \dot1},M_{\dot1}^{\ \dot2},M_{\dot2}^{\ \dot1},M_1^{\ 1},M_1^{\ 2},M_2^{\ 1},P_{\dot1}^{\ 1},P_{\dot1}^{\ 2},P_{\dot2}^{\ 1},P_{\dot2}^{\ 2},K_1^{\ \dot1},K_1^{\ \dot2},K_2^{\ \dot1},K_2^{\ \dot2}\}\ .
\end{equation}
The Killing form is given in terms of the supertrace in the fundamental representation
\begin{equation}
    K^{ab} = \frac14\text{str}(X^a X^b)\ .
\end{equation}
In particular, this fixes the the normalisation of the quadratic Casimir as
\begin{equation}
    C_2 = -\frac{3}{16} R^2 + \frac14 D^2 + ...\ .
\end{equation}

\subsection{Cartan coordinates}

Here we write the relation between primed and unprimed Cartan coordinates. The primed bosonic coordinates are equal to the unprimed, so we only need to consider the fermionic ones. Let $L$ be the matrix
\begin{equation}
    L(\varphi,\theta,\psi) = \begin{pmatrix}
    e^{\frac{i}{2}(\varphi + \psi)}\cos\frac{\theta}{2} & - i e^{\frac{i}{2}(\psi - \varphi)}\sin\frac{\theta}{2}\\
    - i e^{\frac{i}{2}(\varphi - \psi)}\sin\frac{\theta}{2} & e^{-\frac{i}{2}(\varphi + \psi)}\cos\frac{\theta}{2} \end{pmatrix}\ .
\end{equation}
Then the relation between two sets of coordinates reads
\begin{align}
    & \begin{pmatrix}
    q_1'\\
    q_2'
    \end{pmatrix} = e^{\kappa - \frac12\lambda_l}L(\varphi^l_2,\theta^l_2,-\psi^l_1)\begin{pmatrix}
    q_1\\
    q_2
    \end{pmatrix},\quad \begin{pmatrix}
    q'^{\dot1}\\
    q'^{\dot2}
    \end{pmatrix} = e^{\frac12\lambda_r}L(\psi_1^r,-\theta_1^r,\varphi_1^r)\begin{pmatrix}
    q^{\dot1}\\
    q^{\dot2}
    \end{pmatrix},\\
    & \begin{pmatrix}
    s'_{\dot1}\\
    s'_{\dot2}
    \end{pmatrix} = e^{\kappa + \frac12\lambda_l}L(\varphi^l_1,-\theta^l_1,\psi^l_1)\begin{pmatrix}
    s_{\dot1}\\
    s_{\dot2}
    \end{pmatrix},\quad \begin{pmatrix}
    s'^1\\
    s'^2
    \end{pmatrix} = e^{-\frac12\lambda_r}L(\psi_2^r,\theta_2^r,\varphi_2^r)\begin{pmatrix}
    s^1\\
    s^2
    \end{pmatrix}\ .
\end{align}
From these equations, one gets relations between partial derivatives. Among others, we have
\begin{equation}
    \partial_{\kappa'} = \partial_\kappa - q_\alpha \partial_{q_\alpha}, \quad \partial_{\lambda_l'} = \partial_{\lambda_l} + \frac12 q_\alpha\partial_{q_\alpha}, \quad \partial_{\lambda_r'} = \partial_{\lambda_r} - \frac12 q^{\dot\alpha}\partial_{q^{\dot\alpha}},\\
\end{equation}
from which follow covariance conditions in the unprimed coordinates
\begin{equation}
     r = \partial_\kappa -\begin{pmatrix}
    0 & 0 & 0 & 0\\
    0 & 1 & 0 & 0\\
    0 & 0 & 1 & 0\\
    0 & 0 & 0 & 2
    \end{pmatrix}, \quad 2a = \partial_{\lambda_l} + \frac12 \begin{pmatrix}
    0 & 0 & 0 & 0\\
    0 & 1 & 0 & 0\\
    0 & 0 & 1 & 0\\
    0 & 0 & 0 & 2
    \end{pmatrix}, \quad 2b = \partial_{\lambda_r} - \frac12 \begin{pmatrix}
    0 & 0 & 0 & 0\\
    0 & 1 & 0 & 0\\
    0 & 0 & 1 & 0\\
    0 & 0 & 0 & 2
    \end{pmatrix}\ .
\end{equation}
Similarly
\begin{align}
    & 0 = \partial_{\varphi^r_1} + \frac{i}{2}\cos\theta_1^r\left(q^{\dot2}\partial_{q^{\dot2}}-q^{\dot1}\partial_{q^{\dot1}}\right) + \frac12\sin\theta^r_1\left(e^{-i\psi_1^r} q^{\dot2}\partial_{q^{\dot1}} - e^{i\psi_1^r}q^{\dot1}\partial_{q^{\dot2}}\right),\\
    & 0 = \partial_{\psi_1^r} - \frac{i}{2}q^{\dot1}\partial_{q^{\dot1}} + \frac{i}{2}q^{\dot2}\partial_{q^{\dot2}}, \quad 0 = \partial_{\theta^r_1} - \frac{i}{2}e^{-i\psi^r_1} q^{\dot2}\partial_{q^{\dot1}} - \frac{i}{2}e^{i\psi^r_1}q^{\dot1}\partial_{q^{\dot2}},\\
    & 0 = \partial_{\varphi^l_2} - \frac{i}{2} q_1 \partial_{q_1} + \frac{i}{2} q_2 \partial_{q_2}, \quad 0 = \partial_{\theta^l_2} + \frac{i}{2} e^{-i\varphi^l_2}q_2\partial_{q_1} + \frac{i}{2} e^{i\varphi^l_2} q_1\partial_{q_2},\\
    & 0 = \partial_{\psi^l_1} + \frac{i}{2}\cos\theta^l_2 \left(q_1\partial_{q_1} - q_2\partial_{q_2}\right) + \frac12\sin\theta^l_2\left(e^{-i\varphi^l_2} q_2\partial_{q_1} - e^{i\varphi^l_2} q_1\partial_{q_2}\right)\ .
\end{align}

\section{Representations of $\mathfrak{sl}(4|1)$ and $\mathfrak{so}(1,5)$}

Parabolic Verma modules of $\mathfrak{g} = \mathfrak{sl}(4|1)$ are labelled by quantum numbers $(j_1,j_2,\Delta,r)$. Two half-integer spins $(j_1,j_2)$ specify a finite dimensional representation of the rotation Lie algebra $\mathfrak{so}(4) = \mathfrak{su}(2)\oplus\mathfrak{su}(2)$, while $\Delta$ and $r$ are weights of the generators $D$ and $R$, respectively. The value of the quadratic Casimir in the representation $(j_1,j_2,\Delta,r)$ is
\begin{equation}
    C_2(j_1,j_2,\Delta,r) = \frac12 j_1 (j_1 + 1) + \frac12 j_2 (j_2 + 1) + \frac14 \Delta (\Delta - 2) - \frac{3}{16} r^2\ .
\end{equation}
Sometimes, the labels $\Delta,r$ are traded for $q,\bar q$ which are defined by
\begin{equation}
    \Delta = q + \bar q, \quad r = \frac23 (q - \bar q)\ .
\end{equation}
Symmetric traceless representations of $\mathfrak{so}(4)$ have $j_1 = j_2 = l/2$ and the quadratic Casimir is then
\begin{equation}
    C_2(l,\Delta,r) = \frac14 l(l+2) + \frac14\Delta (\Delta - 2) - \frac{3}{16} r^2\ .
\end{equation}
The conformal algebra in four dimensions, $\mathfrak{g}^b = \mathfrak{so}(1,5)$ has highest weight representations labelled by $(j_1,j_2,\Delta)$. The value of the quadratic Casimir in these representations is
\begin{equation}
    C_2(\Delta,j_1,j_2) = \frac12 j_1 (j_1 + 1) + \frac12 j_2 (j_2 + 1) + \frac14\Delta(\Delta - 4)\ .
\end{equation}
Without loss of generality we assume $j_1\leq j_2$ and use alternative notation $l = 2j_1, p = 2(j_2 - j_1)$. Then the Casimir can be rewritten as
\begin{equation}
    C_2(\Delta,l,p) = \frac14 l^2 + \frac18(p+2)(p+2l) + \frac14\Delta(\Delta-4)\ .
\end{equation}
In particular, in the main text we are interested in the cases $p=0$ and $p=1$, which give
\begin{equation}
    C_2(\Delta,l,0) = \frac14 l(l+2) + \frac14 \Delta(\Delta-4), \quad C_2(\Delta,l,1) = \frac14 l(l+3) + \frac14\Delta(\Delta-4) + \frac38\ .
\end{equation}

\section{Bosonic conformal blocks}

In this appendix, we write the scalar and seed blocks that are used throughout the main text. We define
\begin{equation}
    \mathcal{F}^{-(a,b;c)}_{\rho_1,\rho_2}(z,\bar z) = z^{\rho_1}\ _2F_1(a+\rho_1,b+\rho_1;c+2\rho_1;z)\ \bar z^{\rho_2}\ _2F_1(a+\rho_2,b+\rho_2;c+2\rho_2;\bar z) - (z \leftrightarrow \bar z)\ .
\end{equation}
Scalar blocks in the Calogero-Sutherland gauge read
\begin{equation}
    \phi^{a,b}_{\Delta,l}(z,\bar z) = \frac{((z-1)(\bar z-1))^{\frac{a+b}{2}+\frac14}}{(z \bar z)^\frac12} \mathcal{F}^{-(a,b;0)}_{\frac{\Delta+l}{2},\frac{\Delta-l-2}{2}}(z,\bar z)\ .
\end{equation}
Let $G = (G_0^{(1)},G_1^{(1)})^T$ and $\bar G = (\bar G_0^{(1)},\bar G_1^{(1)})$ be the seed and conjugate seed conformal blocks from \cite{Echeverri:2016dun} and set
\begin{equation}
    \Psi^{a,b}_{\Delta,l} = \frac{1}{\sqrt{2}}\begin{pmatrix}
    1 & -1\\
    1 & 1
    \end{pmatrix} S^{-1} G,\quad \bar\Psi^{a,b}_{\Delta,l} = \frac{1}{\sqrt{2}}\begin{pmatrix}
    1 & -1\\
    1 & 1
    \end{pmatrix} S^{-1} \bar G\ .
\end{equation}
The matrix $S$ is defined in the appendix A of \cite{Schomerus:2017eny}, where in the notation of that paper $x = u_1$ and $y = u_2$.

\end{document}